\documentclass[12pt,preprint]{aastex}

\slugcomment{Prepared for ApJ}
\shorttitle{}
\shortauthors{}

\begin{document}

\title{Dynamics and Transit Variations of Resonant Exoplanets}

\author{David Nesvorn\'y}
\affil{Department of Space Studies, Southwest Research Institute,
      1050 Walnut St., Ste 300, \\ Boulder, CO 80302,
      E-mail: davidn@boulder.swri.edu}
\author{David Vokrouhlick\'y}
\affil{Institute of Astronomy, Charles University,
       V Hole\v{s}ovi\v{c}k\'ach 2, CZ--18000 Prague 8, \\
       Czech Republic, E-mail: vokrouhl@cesnet.cz}

\begin{abstract}
The Transit Timing Variations (TTVs) are deviations of the measured mid-transit times from the exact periodicity. 
One of the most interesting causes of TTVs is the gravitational interaction between planets. Here we consider a 
case of two planets in a mean motion resonance (orbital periods in a ratio of small integers). This case is important 
because the resonant interaction can amplify the TTV effect and allow planets to be detected more easily. We develop 
an analytic model of the resonant dynamics valid for small orbital eccentricities and use it to derive the principal 
TTV terms. We find that a resonant system should show TTV terms with two basic periods (and their harmonics). The 
resonant TTV period is proportional $(m/M_*)^{-2/3}$, where $m$ and $M_*$ are the planetary and 
stellar masses. For $m=10^{-4} M_*$, for example, the TTV period exceeds the orbital period by $\sim$2 orders 
of magnitude. The amplitude of the resonant TTV terms scales linearly with the libration amplitude. 
The ratio of the TTV amplitudes of two resonant planets is inversely proportional to the ratio of their masses.
These and other relationships discussed in the main text can be used to aid the interpretation of TTV observations. 
\end{abstract}


\section{Introduction}
Photometric observation of transits is one of the most powerful methods of planet detection. This method 
relies on a possibility that, if the planet's orbit is viewed nearly edge-on, the planet may repeatedly 
transit over the disk of its host star and periodically block a small fraction of the starlight. Thus, by 
monitoring the host star's brightness, the planet's presence can be revealed by a small dip in the 
photometric lightcurve. The main properties of the planet, such as its physical radius and orbital period, 
can be inferred from transit observations. 

The spacing of transit lightcurves would be exactly the same over the course of observations if a planet 
moved on a strictly Keplerian orbit. Several dynamical effects, however, can produce deviations from the 
Keplerian case and induce the Transit Timing Variations (TTVs). TTVs were originally proposed as a non-transiting 
planet detection method (Miralda-Escud\'e 2002, Agol et al. 2005, Holman \& Murray 2005), but have found 
more use in validating the transiting planet candidates from NASA's {\it Kepler} mission (e.g., Holman et al. 
2010; Lissauer et al. 2011). Only a handful of non-transiting planets have been so far detected and 
characterized from TTVs (e.g., Nesvorn\'y et al. 2012, 2013).

A significant progress has been made in the theoretical understanding of various dynamical causes of 
TTVs. These efforts were pioneered by Agol et al. (2005). Heyl \& Gladman (2007) focused on a 
{\it long-period} interaction between planets and showed that the apsidal precession of their orbits can be 
detected only with a long TTV baseline. Nesvorn\'y \& Morbidelli (2008), on the other hand, developed 
a general analytic model for {\it short-period} TTVs and showed that they can be used, under ideal 
circumstances, to uniquely determine the mass and orbital parameters of the interacting planets. 

Explicit analytic formulas for short-period TTVs are now available for zeroth- (Agol et al. 2005, 
Nesvorn\'y \& Vokrouhlick\'y 2014, Deck \& Agol 2015) and first-order terms (Agol \& Deck 2016) in 
planetary eccentricities. The important case of near-resonant TTVs was highlighted in Lithwick et 
al. (2012) for the first-order resonances and in Deck \& Agol (2016) and Hadden \& Lithwick (2016) for the 
second-order resonances.
The near-resonant TTV signal is a special case of the short-periodic variations when one harmonic 
becomes amplified due to a proximity of the system to a mean motion resonance. Vokrouhlick\'y 
\& Nesvorn\'y (2014) considered a case of co-orbital planets and showed that co-orbital TTVs  
are expected to have a characteristic saw-tooth profile (for horseshoe orbits).

While the analytic works cited above cover a wide range of dynamically plausible planetary configurations,
none of them (except for Agol et al. 2005) considered the case of a fully resonant planetary system, where 
two (or more) orbits are {\it inside} a mean motion resonance (e.g., 2:1, 3:2). This case is important because planetary 
migration in a protoplanetary gas disk should bring early planets into resonances (e.g., Masset \& 
Snellgrove 2001). Indeed, many known planetary systems are consistent with having resonant orbits (e.g., 
Winn \& Fabrycky 2015). This motivates us to consider the resonant case. Our main goal is to understand 
how resonant TTVs arise, and how their period and amplitude scale with planetary parameters.

The resonant case has not received much attention in the TTV literature so far at least in part because the 
resonant interaction of planets is complex and not easily amenable to analytic calculations. The short-periodic 
or near-resonant TTV signals, for example, can be computed with the standard methods of perturbation theory, 
where the unperturbed (Keplerian) motion is inserted in the right-hand side of the dynamical equations, 
and a linear variation of the orbital elements is obtained by quadrature. This method fails for the fully 
resonant orbits mainly because the resonant dynamics is non-linear. 

Here, we take advantage of recent advances in the theoretical understanding of resonant dynamics 
(e.g., Batygin \& Morbidelli 2013a; hereafter BM13) and derive approximate formulas for TTVs of a resonant 
pair of exoplanets. In Section 2, we first give a summary of resonant TTVs. The goal of this section 
is to highlight the main results of this work. Sections 3 and 4 explain how 
we obtained these results. We first show how a fully analytic solution can be obtained for a first-order 
resonance (Section 3). We then proceed to expand the exact solution in the Fourier series and explicitly 
derive the periods and amplitudes of the leading TTV terms (Section 4). The application of these results 
to real planetary systems is left for future work. 
\section{Summary of Resonant TTVs}
Consider a planar system of two planets with masses $m_1$ and $m_2$ orbiting a central star 
with mass $M_*$.  The two planets gravitationally interact to produce TTVs. The TTVs signal of 
the two planets, $\delta t_1$ and $\delta t_2$, can be approximated by 
\begin{equation}
\delta t_j = {1 \over n_j} \left( -\delta \lambda_j + 2 \delta h_j\right) + {\cal O}(e_j) \; ,  
\label{ttv}
\end{equation}
where $n_j$ is the mean orbital frequency of planet $j$, $h_j = e_j \sin \varpi_j$, 
$\lambda_j$ is the mean longitude, $e_j$ is eccentricity, and $\varpi_j$ is the longitude of 
periapsis. Observer is assumed to see the system edge-on and measures angles $\lambda_j$ 
and $\varpi_j$ relative to the line of sight. Expression (\ref{ttv}) is valid for small 
orbital eccentricities and small variations of the orbital elements, $\delta \lambda_j$ 
and $\delta h_j$ (Nesvorn\'y 2009). 

Here we consider a case with two planets in a first-order mean motion resonance 
(such that $n_1/n_2 \simeq k/(k-1)$ with integer $k$), and proceed by calculating the 
variations of orbital elements due to the resonant interaction (Sections 3 and 4). The final 
expressions are given as the Fourier series with harmonics of two basic frequencies. 
As for $\delta \lambda_j$ we have
\begin{eqnarray}
-{1 \over n_1} \delta \lambda_1 & = & { 3(k-1) \over \Lambda_1 \nu }  {P_\tau \over 2 \pi} A_\Psi
\left[ (1+\epsilon) \sin f t - {\epsilon \over 4} \sin 2 f t \right]\; , \nonumber \\
-{1 \over n_2} \delta \lambda_2 & = & -{ 3 k \over \Lambda_2 \nu }  {P_\tau \over 2 \pi} A_\Psi
\left[ (1+\epsilon) \sin f t - {\epsilon \over 4} \sin 2 f t \right]\; .
\label{dl}
\end{eqnarray}
Here, $\nu = (3/2)[(k-1)^2 n_1/\Lambda_1 + k^2 n_2/\Lambda_2]$, $\Lambda_j=m_j\sqrt{GM_* a_j}$,
$a_j$ is the semimajor axis, $G$ is the gravitational constant, 
$P_\tau$ is the period of resonant librations in scaled time units ($P_\tau \sim 2$-4 in the most 
cases of interest; Section 3.7), $A_\Psi$ is the amplitude of the resonant oscillations 
of action $\Psi$ (ranging from zero for an exact resonance to $>$1, where the approximation 
used to derive Eq. (\ref{dl}) starts to break down), $f$ is the frequency of resonant librations, 
and $0 \leq \epsilon < 1$ encapsulates the emergence of higher-order harmonics of $f$.\footnote{
Ideally, it would be useful to give the resonant TTV formulas in terms of the orbital elements, but 
these expressions are excessively complex. Here we therefore opt for expressing TTVs in terms 
of the orbital elements, and $A_\Psi$ and $P_\tau$. The dependence on $A_\Psi$ is linear and 
$P_\tau$ admits only a narrow range of values in the libration zone.}

The resonant frequency $f$ is given by
\begin{equation}
f = (\nu C^2)^{1/3} {2 \pi \over P_\tau}\; ,
\label{f}
\end{equation}
with
\begin{equation}
C = {G m_1 m_2 \over a_2} \sqrt{ {f_1^2 \over \Lambda _1} +{f_2^2 \over \Lambda _2}} \; ,
\end{equation}
where $f_1$ and $f_2$ are the resonant coefficients of the Laplacian expansion of the 
perturbing function (Table 1). The scaling of the resonant period, $P=2 \pi/f$, with different 
planetary parameters is discussed in Section 3.7. For practical reasons, Eq. (\ref{dl}) have been 
truncated at the first order in $\epsilon$. Higher orders in $\epsilon$ and higher harmonics
of the libration frequency can be computed using the methods described in Section 4.

The second term in Eq. (\ref{ttv}) is related to the variation of eccentricities and apsidal 
longitudes of the two planets. In the most basic approximation (Section 4.3), it can be 
written as
\begin{eqnarray}
{2 \over n_1} \delta h_1 & = & {P_1 \eta_1^{-{1 \over 3}} \over \pi \sqrt{\Lambda_1}} 
{A A_\Psi \over \sqrt{A^2 + B^2}}  C_{u,0} \sin(\theta_0 + f_\theta t)\; , \nonumber \\
{2 \over n_2} \delta h_2 & = & {P_2 \eta_1^{-{1 \over 3}} \over \pi \sqrt{\Lambda_2}} 
{B A_\Psi \over \sqrt{A^2 + B^2}}  C_{u,0} \sin(\theta_0 + f_\theta t)\; ,
\label{yys2}
\end{eqnarray} 
where $P_j=2 \pi/n_j$ are the orbital periods, 
$\theta_0$ is the initial value of $\theta=k \lambda_2 - (k-1) \lambda_1$, $f_\theta$ is the 
frequency of $\theta$ defined in Eq.~(\ref{ftheta2}). In the libration regime, $f_\theta \ll f$. 
The resonant TTVs can therefore be understood as a sum of librational variations given in Eq. 
(\ref{dl}) and slower modulation of the TTV signal given in Eq. (\ref{yys2}). In Eq. (\ref{yys2}), 
$C_{u,0}$ is a coefficient of the order of unity (Eq. (\ref{cus}) in Section 4.3). For small
libration amplitudes, $A_\Psi C_{u,0} \simeq \sqrt{2 \Psi_0}$, where $\Psi_0$ is the initial
value of the resonant action $\Psi$ defined in Eqs. (\ref{phis}) and (\ref{psi}). This means
that, unlike in Eq. (\ref{dl}), TTVs from the variation of eccentricities and apsidal longitudes 
do not vanish when $A_\Psi=0$. The dependence of the TTV amplitude on planetary masses arises 
from $A$, $B$ and $\eta_1$ in Eq. (\ref{yys2}), where $A=f_1/\sqrt{\Lambda_1}$, 
$B=f_2/\sqrt{\Lambda_2}$, and $\eta_1=\nu/C$. 

\begin{figure*}
 \epsscale{0.9}
  \plotone{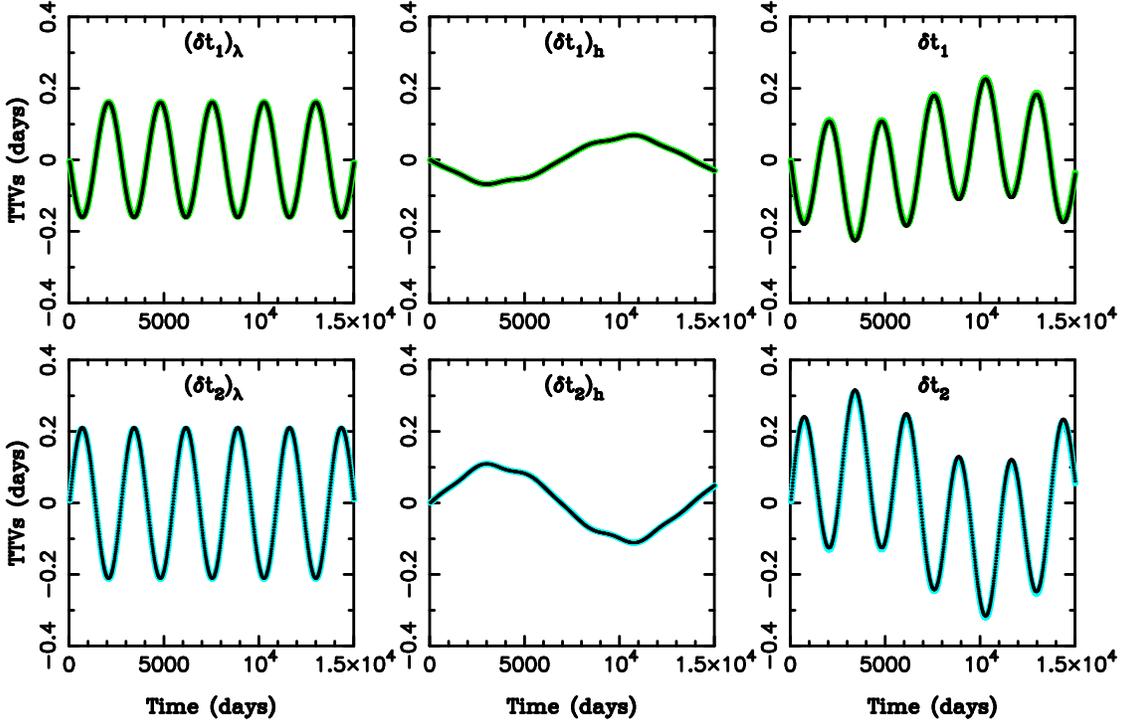}
 \caption{A demonstration of the validity of the analytic TTV formulas obtained in this 
work. Here we set $m_1=m_2=10^{-5} M_*$ with $M_*=M_{\rm Sun}$. The initial orbital elements 
were $a_1=0.1$ AU, $a_2=0.13115$ AU, $e_1=e_2=0.02$, $\lambda_1=\pi$, $\lambda_2=0$, 
$\varpi_1=0$, $\varpi_2=\pi$. This orbital configuration corresponds to the libration regime 
in the 3:2 resonance ($k=3$). The upper (lower) panel shows the results for the inner
(outer) planet. The green and blue lines were computed by numerically integrating the 
differential equations corresponding to the resonant Hamiltonian (\ref{ham3}) 
and (\ref{ham4}). The black lines were obtained from the analytic TTV expressions 
(\ref{ttv}), (\ref{dl}) and a generalization of (\ref{yys2}) derived in Section 4.3.
From left to right the panels show TTVs from $\delta \lambda_j$ and 
$\delta h_j$, and their sum from Eq. (\ref{ttv}). The validity of the analytic model
is excellent in this case because the resonant amplitude is relatively small 
($A_\Psi\simeq0.65$ with $\delta=2.36$ and $\Psi_0=3.01$; see Section 3).}
 \label{ttv1}
\end{figure*}

\begin{figure*}
 \epsscale{0.9}
  \plotone{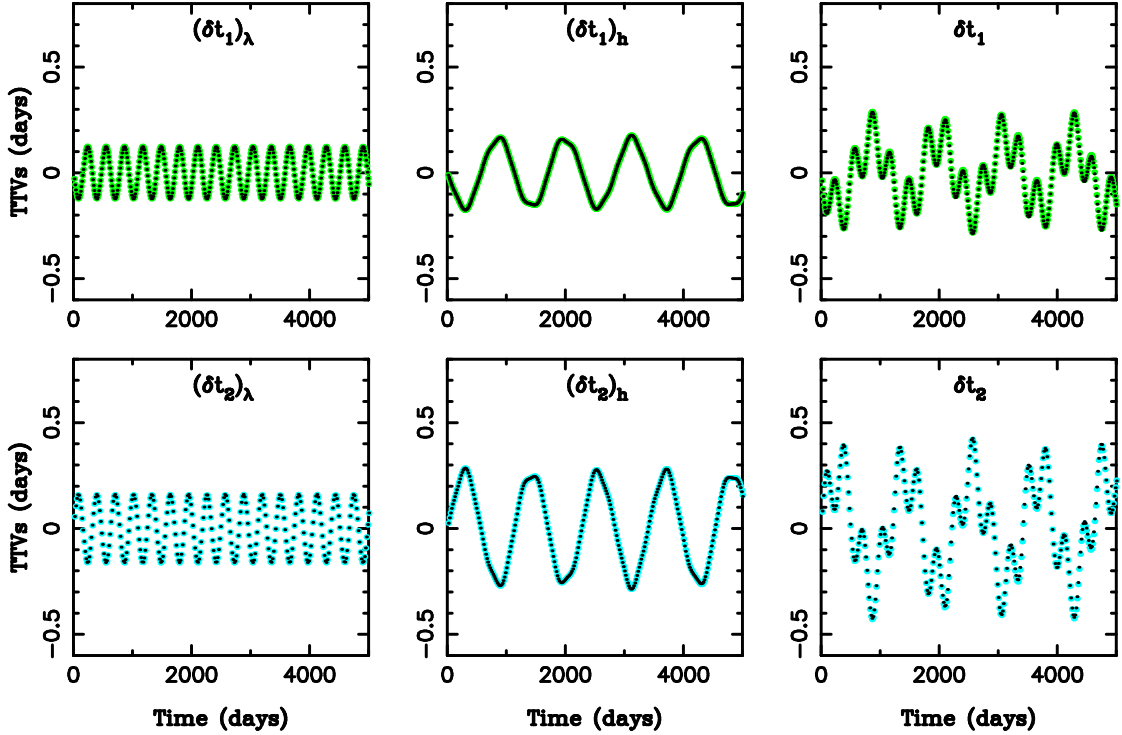}
 \caption{The same as Fig. \ref{ttv1} but with $m_1=m_2=3\times10^{-4} M_*$, 
$M_*=M_{\rm Sun}$, $a_1=0.1$ AU, $a_2=0.132$ AU, $e_1=e_2=0.05$, $\lambda_1=\pi$, 
$\lambda_2=0$, $\varpi_1=0$, $\varpi_2=\pi$. This corresponds to $A_\Psi\simeq-0.65$, 
$\delta\simeq1.38$ and $\Psi_0\simeq1.96$ (Section 3). The periods are shorter in this plot
than in  Fig. \ref{ttv1}, because the two planets were given larger masses.}
\label{ttv2}
\end{figure*}

Two main approximations were adopted to derive Eqs. (\ref{dl}) and (\ref{yys2}). In the first 
approximation, we retained the lowest-order eccentricity terms in the resonant interaction 
of planets. In the second approximation, we assumed that the libration amplitudes are not very large, 
and expanded the exact solution in the Fourier series. Both these approximations are tested in 
Section 5. Here we just illustrate the validity of the Fourier expansion for small libration 
amplitudes (Figures \ref{ttv1} and \ref{ttv2}). 

Let us briefly consider an application of our results to the TTV analysis. We assume that
the photometric transits are detected for planets 1 and 2, and that the orbital periods inferred from 
the transit ephemeris are such that orbital period ratio $P_2/P_1 \simeq k/(k-1)$ with small integer 
$k$ (indicating near-resonant or resonant orbits). Furthermore, TTVs are assumed to be detected for both 
planets. To be specific, let us consider a realistic case with the available TTV data spanning several 
years of observations, which is not long enough to resolve the frequencies related to the apsidal 
precession of orbits. 

The first step of the TTV analysis is to apply the Fourier analysis to the TTV data. This may reveal 
that the TTV signal contains a basic period. In principle, this period can 
be one of following two periods: (1) the super period defined as $P_{\rm s}=(k/P_2-(k-1)/P_1)^{-1}$ 
(e.g, Lithwick et al. 2012), or (2) the resonant libration period $P=2 \pi/f$ with $f$ defined in 
Eq.~(\ref{f}). In the first case, the system is not in the libration regime of the resonance and 
TTVs can therefore be interpreted using the expressions appropriate for the near-resonant dynamics  
(Lithwick et al. 2012, Agol \& Deck 2016). The results described in this work apply in the second case. 
Section 3.7 explains how the libration period can be used to constrain planetary masses. Specifically, 
$P \propto (m/M_*)^{-2/3}$ (Agol et al. 2005, Holman et al. 2010), and therefore larger planetary masses imply shorter TTV 
periods. For the 2:1 resonance with $P_1=10$ days and $m_1\simeq m_2=10^{-4} M_*$, for example, the 
libration period is $P\simeq4.5$ yr (Section 3.7). 

In the next step, it can be useful to check if the Fourier analysis of the TTV data provides evidence 
for harmonics of the basic period. If that's the case, this can indicate 
that the libration amplitude is relatively large. A comparison of the amplitudes of different harmonics 
can then be used to constrain the parameter $\epsilon$ in Eq.~(\ref{dl}), which is related to the 
libration amplitude via equations reported in Appendix A. 

\begin{figure*}
 \epsscale{0.47}
  \plotone{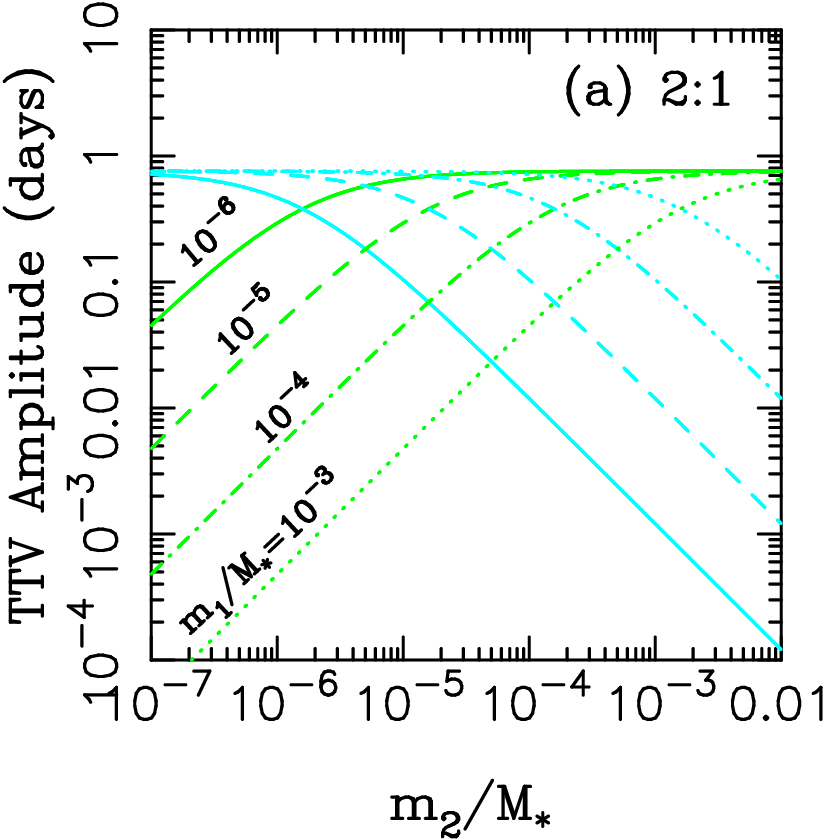}
  \plotone{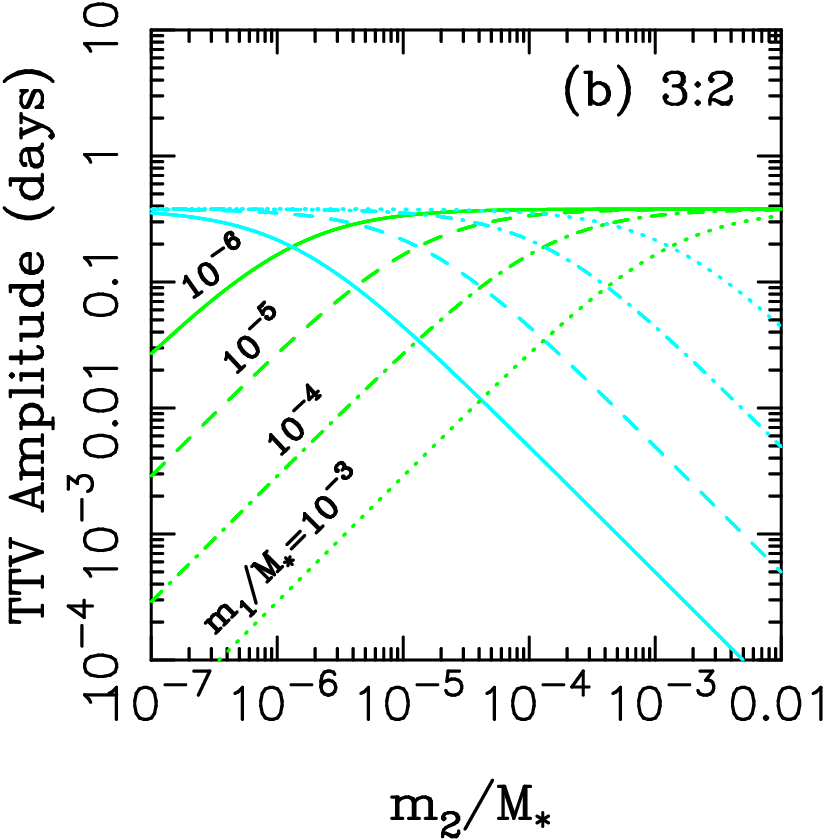}
 \caption{The TTV amplitude from $\delta \lambda_j$ (Eq. \ref{dl}) as a function 
of $m_2/M_*$. Here we assumed that $P_1=10$ days, $P_\tau=3$, $A_\Psi=1$ and computed 
the TTV amplitude for several different values of $m_1/M_*$: $10^{-6}$ (solid lines),
$10^{-5}$ (dashed lines), $10^{-4}$ (dot-dashed lines), $10^{-3}$ (dotted lines).
Panels (a) and (b) show the results for the 2:1 and 3:2 resonances, respectively. 
The green (blue) lines show the amplitude of $\delta t_1$ ($\delta t_2$).}
 \label{ampl2}
\end{figure*}

We then proceed by comparing the TTV amplitudes of the two planets. From Eq. (\ref{dl}) we have
that $\delta \lambda_1/\delta \lambda_2 \simeq -[(k-1)/k]^{2/3} m_2/m_1$. The TTV amplitude ratio therefore
constrains the ratio of planetary masses (Agol et al. 2005). Figure \ref{ampl2} shows how the TTV 
amplitudes depend on planetary masses. For $m_1 \ll m_2$, we obtain from Eq. (\ref{dl})
that the TTV amplitudes of the inner and outer planets, $A_1$ and $A_2$, are
\begin{eqnarray}
A_1 & = & {P_1 \over 2 \pi} {P_\tau \over \pi} {A_\Psi \over k-1}\; , \nonumber \\
A_2 & = & {P_2 \over 2 \pi} {P_\tau \over \pi} {\sqrt{\alpha} A_\Psi \over k-1} {m_1 \over m_2} \; ,
\end{eqnarray}
where we denoted $\alpha=a_1/a_2$. This means that the TTV amplitude of the inner planet
is independent of masses, while that of the outer planet constrains $m_1/m_2$. For $P_1=10$ day,
$P_\tau=3$ and $A_\Psi=1$, $A_1 = 1.52/(k-1)$ days for a $k$:($k-1$) resonance. A similar analysis can 
be performed for $m_1 \gg m_2$. 

In addition to the dependence on masses,
the TTV amplitudes (linearly) depend on the libration amplitude $A_\Psi$. We therefore expect
that some degeneracy should exist between the planetary mass and libration amplitude, with 
large TTV amplitudes being produced either by large masses or large libration amplitudes.
This degeneracy can be broken if the libration frequency harmonics are detected, providing
constraints on the libration amplitude, or if the measured TTV signal also contains 
short-period (chopping) terms (e.g., Nesvorn\'y et al. 2013). A detailed analysis of this problem 
is left for future work. 
\section{Analytic Model of Resonant Dynamics}
Here we discuss a Hamiltonian model of the resonant dynamics. Our approach closely follows 
the work of BM13. We take several shortcuts to simplify the reduction of the Hamiltonian to 
an integrable system. Then, in Section 3.6, we present an exact analytic solution. This 
solution is used in Section 4 to derive Eq. (\ref{dl}) and (\ref{yys2}). 
\subsection{Hamiltonian Formulation of the Problem}
The Poincar\'e canonical variables of two planets orbiting their host star are denoted 
by (${\bf r}_0,{\bf r}_1,{\bf r}_2;{\bf p}_0,{\bf p}_1,{\bf p}_2$). The 
coordinate vector ${\bf r}_0$ defines the host star's position with respect to the 
system's barycenter. Vectors ${\bf r}_1$ and ${\bf r}_2$ are the position vectors  
of the two planets relative to their host star. Momentum ${\bf p}_0$ is the
total linear momentum of the system (${\bf p}_0=0$ in the barycentric inertial 
frame). Momenta ${\bf p}_1$ and ${\bf p}_2$ are the linear momenta of the two 
planets in the barycentric inertial frame. The Poincar\'e variables are canonical, which 
can be demonstrated by calculating their Poisson brackets.

Using Poincar\'e variables the differential equations governing dynamics of 
the two planets can be conveniently written in a Hamiltonian form, where the 
total Hamiltonian is a sum of the Keplerian and perturbation parts, 
${\cal H}= {\cal H}_{\rm K}+{\cal H}_{\rm per}$, with
\begin{equation}
 {\cal H}_{\rm K} = \sum_{j=1}^2 \left(\frac{p_j^2}{2\,
  \mu_j}-G\,\frac{\mu_j M_j}{r_j}\right)\; ,
 \label{ham1}
\end{equation}
and
\begin{equation}
 {\cal H}_{\rm per} = \frac{{\bf p}_1\cdot{\bf p}_2}{M_*}-
  G\,\frac{m_1 m_2}{\left|{\bf r}_1-{\bf r}_2\right|}\; .
  \label{ham2}
\end{equation}
Here we denoted $M_j=m_j+M_*$ and the reduced masses $\mu_j=m_jM_*/M_j$, where $j=1$ and 
2 stand for the inner and outer planet, respectively.

We assume that the planets are near or in a first-order mean motion resonance such that 
the ratio of their orbital periods is $P_2/P_1 \simeq k/(k-1)$ for some integer 
$k\geq 2$. In terms of the osculating orbital elements we have
\begin{equation}
 {\cal H}_{\rm K} = -G\,\frac{\mu_1 M_1}{2a_1}-
  G\,\frac{\mu_2 M_2}{2a_2}\; , \label{ham3}
\end{equation}
where $a_1$ and $a_2$ are the semimajor axes of planets, and
\begin{eqnarray}
 {\cal H}_{\rm per} &\!\!\!=\!\!\!& -G\,\frac{m_1 m_2}{a_2}\times
\nonumber \\
 & & \quad \Bigl\{f_1\,e_1\cos\left[k\lambda_2-\left(k-1\right)\lambda_1-
  \varpi_1\right] \nonumber \\
  & & \quad + f_2\,e_2 \cos\left[k\lambda_2-\left(k-1\right)\lambda_1-
  \varpi_2\right]\Bigr\}\; , \label{ham4}
\end{eqnarray}
where $e_1$ and $e_2$ are the orbital eccentricities, $\lambda_1$ and
$\lambda_2$ are the mean longitudes, and $\varpi_1$ and $\varpi_2$
are the longitudes of pericenter. In Eq. (\ref{ham4}) we only retained 
two most important resonant terms, and the lowest (first-order) eccentricity power. 
This expression is thus valid only near a specific resonance and for 
low orbital eccentricities of both planets. The planetary orbits 
are assumed to be in the same plane such that all inclination-dependent 
terms vanish.

The coefficients $f_1$ and $f_2$ are functions of the semimajor axis ratio 
$\alpha=a_1/a_2<1$, and can be written as 
\begin{eqnarray}
 f_1 & \!\!\!=\!\!\! & -k\, b^{(k)}_{1/2}(\alpha) - \frac{\alpha}{2}
  D b^{(k)}_{1/2}(\alpha) \; , \nonumber \\
 f_2 & \!\!\!=\!\!\! & \left(k-\frac{1}{2}\right) b^{(k-1)}_{1/2}(\alpha)
  +\frac{\alpha}{2} D b^{(k-1)}_{1/2}(\alpha) - \frac{\delta_{k,2}}{\alpha^{1/2}}\; , \label{ham5b} 
\end{eqnarray}
where $b^{(k)}_{1/2}(\alpha)$ are the Laplace coefficients, $D=d/d\alpha$, and
$\delta_{k,2}$ is the Kronecker symbol (e.g., Brouwer \& Clemence 1961).
The last term in the expression for $f_2$ only appears if $k=2$ (2:1 resonance). 
The Laplace coefficients were computed using the recurrences described in Brouwer 
\& Clemence (1961, Secs.~15.7 and 15.8). Their values are reported in Table 1.
\subsection{Resonant Variables}
The resonant Hamiltonian discussed in the previous section needs to be written 
in canonical variables. The standard choice is the Delaunay elements
\begin{eqnarray}
\Lambda_j & \!\!\!=\!\!\! & \mu_j \sqrt{G M_j a_j} \; , \quad
\lambda_j \; , \nonumber \\
\Gamma_j & \!\!\!=\!\!\! & \Lambda_j\left(1-\sqrt{1-e_j^2}\right) \; , \quad
\gamma_j = -\varpi_j   \; . 
\end{eqnarray}
Note that $\Gamma_j \simeq (1/2)\Lambda_j e_j^2$ for small $e_j$. The elements 
$(\Lambda_j$,$\Gamma_j)$ are the canonical momenta, and ($\lambda_j,\gamma_j$) are 
the conjugated canonical coordinates.     

The resonant Hamiltonian ${\cal H}$ written in terms of the Delaunay elements has  
four degrees of freedom (DOF). Using canonical transformations we will reduce it 
to an integrable one-DOF system. We first perform a canonical transformation to 
the resonant canonical variables defined as
\begin{eqnarray}
 K_1 & = & \Lambda_1+(k-1)(\Gamma_1+\Gamma_2) \; , \quad \lambda_1 \; , \nonumber \\
 K_2 & = & \Lambda_2-k(\Gamma_1+\Gamma_2) \; , \quad \quad \quad \; \;  \lambda_2 \; , \nonumber \\
 \Gamma_1 &  &  , \quad \sigma_1  =  k\lambda_2-\left(k-1\right)\lambda_1- \varpi_1   \; , \nonumber \\
 \Gamma_2 &  &  , \quad \sigma_2  =  k\lambda_2-\left(k-1\right)\lambda_1- \varpi_2   \; .
\label{ham6b} 
\end{eqnarray}
When the new variables are inserted in Eqs. (\ref{ham3}) and (\ref{ham4}), it becomes clear
that the resonant Hamiltonian depends on $\sigma_1$ and $\sigma_2$, but not on $\lambda_1$ 
and $\lambda_2$. Therefore, ${\rm d}K_j/{\rm d}t=-\partial{\cal H}/\partial\lambda_j=0$ and
both momenta $K_1$ and $K_2$ are the new constants of motion. By summing them, we find that
$K_1+K_2=\Lambda_1+\Lambda_2-\Gamma_1-\Gamma_2$ is the total angular momentum of the system.
Also, $k K_1 + (k-1) K_2 = k\Lambda_1+(k-1)\Lambda_2={\rm const.}$ implies that any small 
changes of the semimajor axes, $\delta a_1$ and $\delta a_2$, must be anti-correlated, and 
have relative amplitudes such that $\delta a_1/\delta a_2 \simeq - \alpha_{\rm res}^{1/2} (k-1) 
m_2 / k m_1$, where we denoted $\alpha_{\rm res}=[(k-1)/k]^{2/3}$.
\subsection{Approximation for Small Semimajor Axis Changes}
The transformation to the resonant canonical variables produced a 2-DOF Hamiltonian ${\cal H}=
{\cal H}(\sigma_1,\sigma_2;\Gamma_1,\Gamma_2)$. In the next step, we assume that any changes 
of the semimajor axes of planets are small. Specifically, we write 
$a_j=a_j^*+\delta a_j$, where $a_j^*$ is some reference value and $\delta a_j \ll a_j^*$,
insert this expression into the Keplerian part of the Hamiltonian (Eq.~\ref{ham3}), and
expand it in powers of $\delta a_j$. The first and second-order terms in $\delta a_j$ 
are retained. We then use $\delta a_j/a_j^*=2 \delta \Lambda_j/\Lambda_j^*$ and rewrite
all expressions in terms of $\Lambda_j^*$ and $\delta \Lambda_j$.\footnote{A general result 
can also be obtained by directly performing the Taylor expansion in $\Lambda_j=\Lambda_j^*+\delta \Lambda_j$.} 
Finally, we substitute $\delta \Lambda_j \rightarrow \Lambda_j-\Lambda_j^*$ and drop all (dynamically 
unimportant) constant terms. This substitution is useful because it allows us to work with 
$\Lambda_1$ and $\Lambda_2$ instead of their variations $\delta \Lambda_1$ and $\delta \Lambda_2$. 
Finally, we express ${\cal H}_{\rm K}$ in terms of the canonical variables defined in 
Eq. (\ref{ham6b}). This leads to  
\begin{equation}
{\cal H}_{\rm K}  = n_0 +  n_{\rm s} (\Gamma_1 + \Gamma_2)
                -  \nu (\Gamma_1 + \Gamma_2)^2\; 
\label{hk}  
\end{equation}
with
\begin{eqnarray}
n_0 & = & 4(n_1 K_1 + n_2 K_2) - {3 \over 2} \left[ n_1 {K_1^2 \over \Lambda_1^*} + 
n_2 {K_2^2 \over \Lambda_2^*} \right]\; , \nonumber \\
n_{\rm s} & = & 4[k n_2 - (k-1) n_1]-3\left[ k K_2 {n_2 \over \Lambda_2^*} - 
(k-1) K_1 {n_1 \over \Lambda_1^*} \right]\; , \nonumber \\
\nu & = & {3 \over 2} \left[  (k-1)^2 {n_1 \over \Lambda_1^*} 
+  k^2 {n_2 \over \Lambda_2^*} \right]\; .
\label{super}
\end{eqnarray}
Here we denoted $n_j=\sqrt{G M_j/a_j^{*3}}$ and $\Lambda_j^*=\mu_j\sqrt{G M_j a_j^*}$. The quantities 
$n_j$, $n_{\rm s}$ and $\nu$ are constant parameters.
Note that $n_{\rm s}$ is related to the so-called {\it super} frequency, which is the expected frequency 
of the TTV signal for a pair of near-resonant planets (e.g., Lithwick et al. 2012). In addition to
the usual term, $k n_2 - (k-1) n_1$, here $n_{\rm s}$ also includes a correction that is a second order 
in the eccentricity (through its dependence on $K_1$ and $K_2$). 

As for ${\cal H}_{\rm per}$ in Eq. (\ref{ham4}), we have that $e_j=\sqrt{2 \Gamma_j/\Lambda_j}$
for small eccentricity. In addition, because ${\cal H}_{\rm per}$ is already small, we do not 
need to retain terms proportional to $\delta a_j$ (thus $\Lambda_j \rightarrow \Lambda_j^*$).
The perturbation function then admits the following form
\begin{equation}
 {\cal H}_{\rm per} =  -G\,\frac{m_1 m_2}{a_2^*}
\left[A \sqrt{2 \Gamma_1} \cos \sigma_1 + B \sqrt{2 \Gamma_2} \cos \sigma_2\right]\, ,
\label{hper}
\end{equation}
where we introduced constants $A=f_1/\sqrt{\Lambda_1^*}$ and $B=f_2/\sqrt{\Lambda_2^*}$. 
\subsection{Reducing Transformation}
The Hamiltonian in Eqs. (\ref{hk}) and (\ref{hper}) has 2 DOF. It can be reduced to 
1 DOF by the following canonical transformations (Sessin \& Ferraz-Mello 1984, Wisdom 1986, 
Henrard et al. 1986). First, we move from variables $(\sigma_1,\sigma_2;\Gamma_1,\Gamma_2)$ 
to $(y_1,y_2;x_1,x_2)$ such that
\begin{eqnarray}
  x_1 & \!\!\!=\!\!\! & \sqrt{2\Gamma_1}\cos\sigma_1\; , \quad
   y_1= \sqrt{2\Gamma_1}\sin\sigma_1 \; , \label{ham10a} \nonumber \\
  x_2 & \!\!\!=\!\!\! & \sqrt{2\Gamma_2}\cos\sigma_2\; , \quad
   y_2= \sqrt{2\Gamma_2}\sin\sigma_2 \; . \label{ham10b} 
\end{eqnarray}
Second, we perform a reducing transformation to the new variables $(v_1,v_2;u_1,u_2)$ 
defined as
\begin{eqnarray}
  u_1 & \!\!\!=\!\!\! & \frac{Ax_1+Bx_2}{\sqrt{A^2+B^2}}\; , \quad
   v_1= \frac{Ay_1+By_2}{\sqrt{A^2+B^2}} \; , \label{ham11a} \nonumber \\
  u_2 & \!\!\!=\!\!\! & \frac{Bx_1-Ax_2}{\sqrt{A^2+B^2}}\; , \quad
   v_2= \frac{By_1-Ay_2}{\sqrt{A^2+B^2}} \; . \label{redt}
\end{eqnarray}
And last, we introduce new polar variables $(\phi_1,\phi_2;\Phi_1,\Phi_2)$ such that
\begin{eqnarray}
  u_1 & \!\!\!=\!\!\! & \sqrt{2\Phi_1}\cos\phi_1\; , \quad
   v_1= \sqrt{2\Phi_1}\sin\phi_1 \; , \nonumber \\
  u_2 & \!\!\!=\!\!\! & \sqrt{2\Phi_2}\cos\phi_2\; , \quad
   v_2= \sqrt{2\Phi_2}\sin\phi_2 \; . \label{ham14b} 
\end{eqnarray}
It can be shown that $\Gamma_1+\Gamma_2=\Phi_1+\Phi_2$.
Therefore, after dropping the first constant term in Eq. (\ref{hk}), and rewriting Eq. 
(\ref{hper}) in the new variables, the Hamiltonian becomes
\begin{equation}
 {\cal H} = n_{\rm s}(\Phi_1+\Phi_2)-\nu(\Phi_1+\Phi_2)^2-C\sqrt{2\Phi_1}\cos\phi_1\; ,
\label{hnew}
\end{equation}
where we denoted $C=G m_1 m_2 \sqrt{A^2+B^2}/a^*_2$. Notably, the new Hamiltonian 
(\ref{hnew}) is independent of $\phi_2$, and the canonical momentum $\Phi_2$ is therefore 
a new constant of motion. That's the magic of the reducing transformation. 

The momenta $\Phi_1$ and $\Phi_2$ defined by the transformations discussed above can be
expressed in a compact form
\begin{eqnarray}
\Phi_1 & = & {1 \over 2} {|A z_1 + B z_2|^2 \over A^2+B^2} \; , \nonumber\\
\Phi_2 & = & {1 \over 2} {|B z_1 - A z_2|^2 \over A^2+B^2} \; , 
\label{phis}
\end{eqnarray}
where $z_j=x_j+\imath y_j = \sqrt{2 \Gamma_j} \exp \imath \sigma_j$. Condition
$\Phi_2={\rm const.}$ thus defines a circle in the plane of complex variables $z_1$ 
and $z_2$ and requires that $B z_1 - A z_2$ lies on the circle at any  
time. Also, $\phi_1 = {\rm arg}(A z_1 + B z_2)$ and $\phi_2 = {\rm arg}(B z_1 - 
A z_2)$, where ${\rm arg}(z)$ denotes the argument of~$z$.  
\subsection{Final Scaling}
The Hamiltonian (\ref{hnew}) depends on parameters $n_{\rm s}$, $\nu$ and $C$. We rescale 
$\Phi_1$ and time to bring the Hamiltonian to a simple form
\begin{equation}
{\cal H}=-(\Psi-\delta)^2-\sqrt{2 \Psi} \cos \psi\; ,
\label{fm}
\end{equation}
where the parametric dependence is expressed by a single parameter 
\begin{equation}
\delta  =  \eta_1^{2/3} \left[ {n_{\rm s} \over 2 \nu} - \Phi_2 \right]\; 
\label{delta}
\end{equation}
with $\eta_1=\nu/C$. Here, $\psi=\phi_1$, and
\begin{equation}
\Psi   =  \eta_1^{2/3} \Phi_1\; .
\label{psi}
\end{equation}
The Hamiltonian equations are 
\begin{eqnarray}
{ {\rm d}\psi \over {\rm d}\tau} & = & {\partial {\cal H} \over \partial \Psi} = 
-2(\Psi-\delta)-{1 \over \sqrt{2 \Psi}} \cos \psi\; , \label{hameq1}\\
{ {\rm d}\Psi \over {\rm d}\tau} & = & -{\partial {\cal H} \over \partial \psi} = 
-\sqrt{2 \Psi} \sin \psi\; ,
\label{hameq2}
\end{eqnarray}  
where $\tau$ relates to normal time $t$ by
\begin{equation}
\tau = \eta_2^{1/3} t\;  
\label{tscale}
\end{equation}
with $\eta_2=\nu C^2$. When a solution of Eqs. (\ref{hameq1}) and (\ref{hameq2}) is found, the scaling 
parameters $\eta_1$ and $\eta_2$ can be used to map the solution back to the original variables.

The Hamiltonian (\ref{fm}) and the corresponding Eqs. (\ref{hameq1}) and (\ref{hameq2}) have 
been extensively studied in the past. They are equivalent to the second fundamental model of resonance 
(Henrard \& Lema\^{\i}tre 1983), and to the Andoyer model discussed in Ferraz-Mello (2007). Here we 
first consider the dynamical flow arising from this Hamiltonian and its dependence on $\delta$. 
In the next section, we show that Eqs. (\ref{hameq1}) and (\ref{hameq2}) have an exact analytic solution
in terms of the Weierstrass elliptic functions.

\begin{figure*}
 \epsscale{0.7}
  \plotone{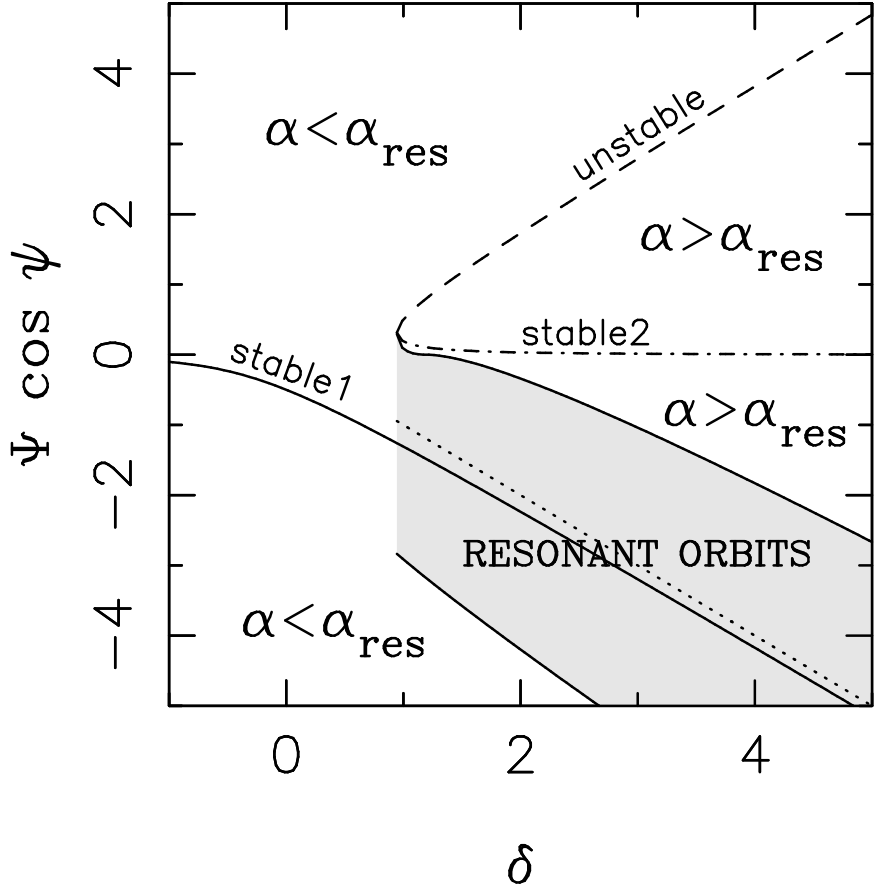}
 \caption{The equilibrium points and different dynamical regimes of the resonant 
Hamiltonian (\ref{fm}). The solid line denoted `stable 1' is the first stable 
equilibrium that exists for any value of $\delta$. The second stable equilibrium,
denoted by `stable 2', appears only for $\delta>\delta_*\simeq0.945$. The dashed 
line is the unstable equilibrium. The gray area is the place where the resonant 
librations occur. The dynamical regime where the two orbits are just wide (narrow) 
of the resonance is labeled by $\alpha<\alpha_{\rm res}$ ($\alpha>\alpha_{\rm res}$).
The dotted line is an approximation of the first equilibrium point, $\Psi_{\rm eq}=\delta$,
that becomes progressively better with increasing $\delta$.}
 \label{delta2}
\end{figure*}

\begin{figure*}
 \epsscale{0.444}
  \plotone{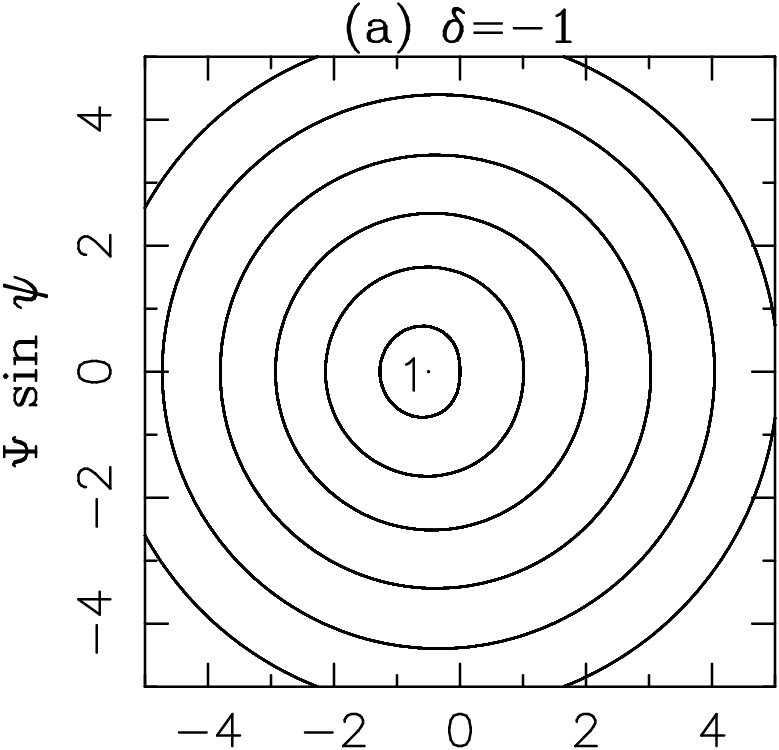}
 \epsscale{0.4}
  \plotone{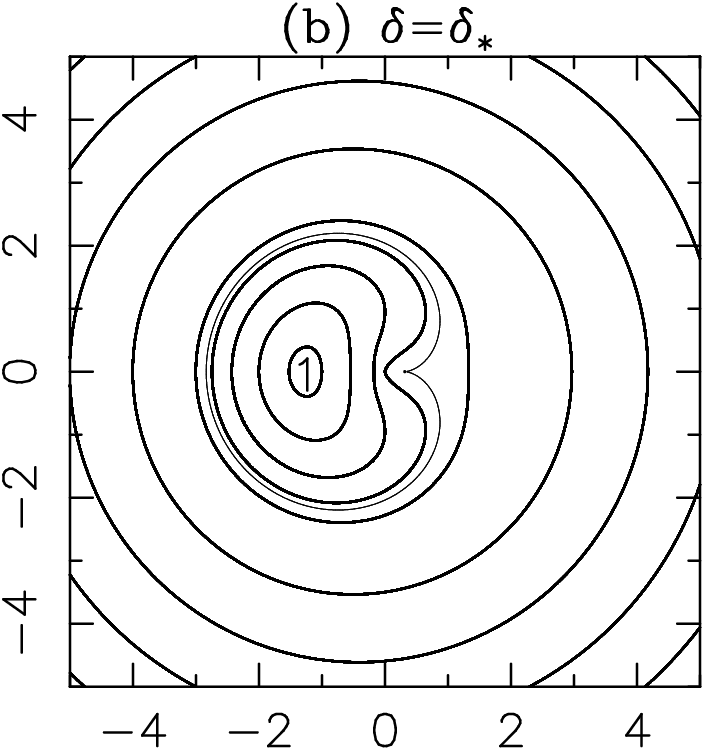}
\vspace*{2.mm}
 \epsscale{0.444}
  \plotone{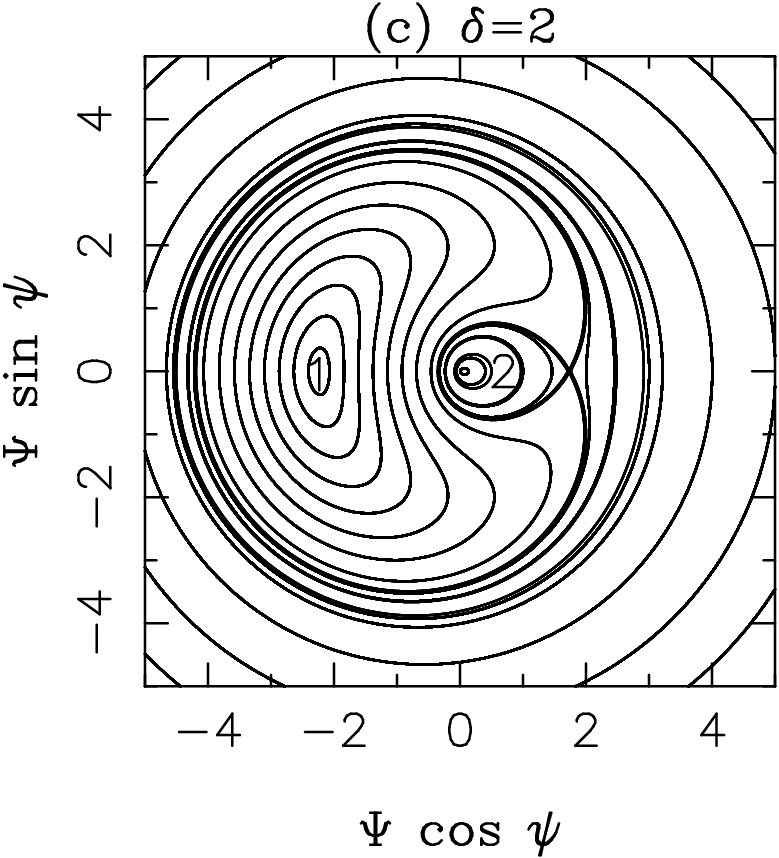}
 \epsscale{0.4}
  \plotone{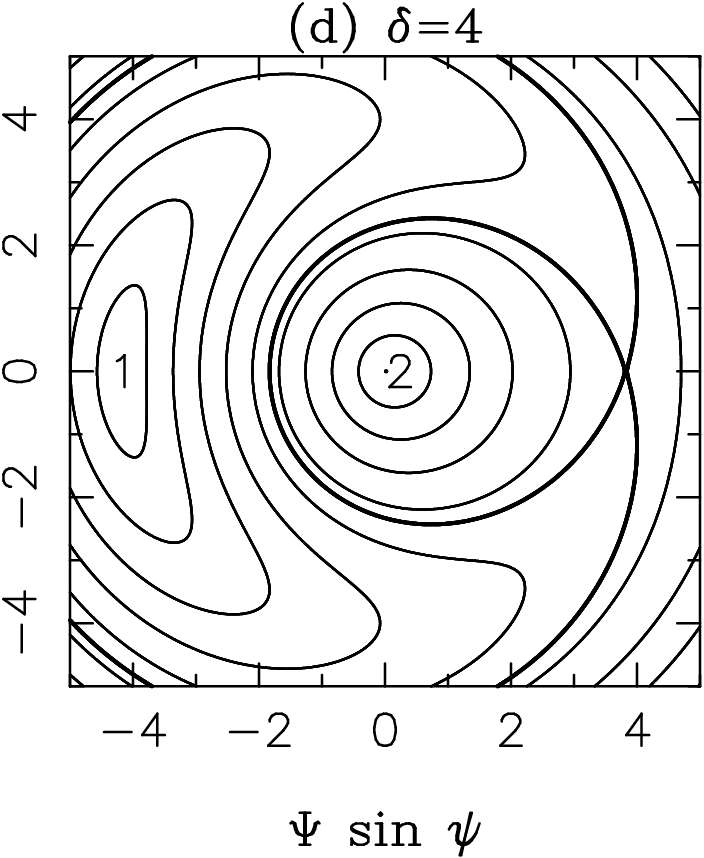}
 \caption{Dynamical portraits for four different values of parameter $\delta$: 
(a) $\delta=-1$, (b) $\delta=\delta_*$, (c) $\delta=2$, and (d) $\delta=4$.
The two stable equilibria are labeled `1' and `2'. The cusp trajectory is shown 
by a thin line in panel (b). The separatrices are shown by bold lines in panels (c) and (d).}  
 \label{portraits}
\end{figure*}

The equilibrium points of Eqs. (\ref{hameq1}) and (\ref{hameq2}) control the general structure of the dynamical flow.
Since ${\rm d}\Psi/{\rm d}\tau=0$ implies that $\sin \psi=0$ in (\ref{hameq2}), the equilibrium points occur for 
$\psi=0$ or $\pi$. The equilibrium values of $\Psi$ are obtained from ${\rm d}\psi/{\rm d}\tau=0$, 
which leads to a problem of finding roots of a  cubic equation $\Psi^3-2\delta\Psi^2+\delta^2\Psi-1/8=0$. 
There is only one (stable) equilibrium point for $\delta\leq\delta_*=\left(27/32\right)^{1/3}\simeq0.945$. This 
equilibrium point is located at $\psi=\pi$ and $0<\Psi<1.26$ (Figure \ref{delta2}). The dynamical 
flow around the equilibrium point is simple. When projected
to the $(\Psi \cos \psi, \Psi \sin \psi)$ plane,\footnote{It is more common in this context 
to use $(\sqrt{2 \Psi} \cos \psi, \sqrt{2 \Psi} \sin \psi)$, because these variables are 
canonical.} the trajectories are concentric, slightly deformed circles centered on the
equilibrium point (Figure \ref{portraits}a). $\Psi$ changes only slightly during each cycle, 
and $\psi$ circulates in the clockwise direction for most initial conditions except for the ones 
located very close to the equilibrium point, where $\psi$ oscillates around $\pi$. 

For $\delta=\delta_*$, the stable equilibrium point is already substantially displaced 
from the origin. The dynamical transition is heralded by the appearance of a cusp trajectory 
(shown by a thin line in Figure \ref{portraits}b). Three equilibrium points appear for 
$\delta>\delta_*$, two of which are stable and one is unstable. The unstable equilibrium 
point is traversed by a separatrix (shown by thick lines in Figures \ref{portraits}c and \ref{portraits}d),
which surrounds the truly resonant trajectories for which $\psi$ librates around $\pi$. The 
stable equilibrium point in the center of the resonant island is a smooth extension of the 
equilibrium point from $\delta < \delta_*$ to $\delta > \delta_*$. It is located 
at $\psi=\pi$ and $\Psi=\Psi_{\rm eq}$, where $\Psi_{\rm eq}$ increases with $\delta$
(Figure \ref{delta2}). For large $\delta$, $\Psi_{\rm eq} \simeq \delta$ (Henrard \& Lema\^{\i}tre 
1983). 
\subsection{Exact Analytic Solution}
A rearrangement of Eqs. (\ref{hameq1}) and (\ref{hameq2}) shows that the momentum $\Psi$ 
satisfies
\begin{equation}
 \left(\frac{{\rm d}\Psi}{{\rm d} \tau }\right)^2 = f\left(\Psi\right)\; , \label{sfmr2}
\end{equation}
with the right-hand side being a quartic polynomial
\begin{equation}
 f\left(\Psi\right) = a_0\Psi^4+4 a_1\Psi^3+6a_2\Psi^2+4a_3\Psi + a_4\; .
  \label{sfmr3}
\end{equation}
The coefficients are 
\begin{eqnarray}
 a_0 & = & -1 \; , \nonumber \\
 a_1 & = & \phantom{-}\delta \; , \nonumber \\
 a_2 & = & -\frac{1}{3}\left[h_0+3\delta^2\right] \; , \nonumber \\
 a_3 & = & \phantom{-}\frac{1}{2}\left[1+2\delta\left(h_0+\delta^2
  \right)\right] \; , \nonumber \\
 a_4 & = & -\left[h_0+\delta^2\right]^2 \; , \label{sfmr4e} 
\end{eqnarray}
where ${\cal H}\left(\Psi_0,\psi_0\right)=h_0$ is the conserved energy,
and $(\Psi_0,\psi_0)$ is the initial condition. Equation~(\ref{sfmr2}) admits a general 
analytic solution (see Whittaker \& Watson 1920, Sec.~20.4)
\begin{equation}
 \Psi(\tau) =\Psi_0 + \frac{{\cal S}\sqrt{f_0}\,\wp^\prime\left(\tau-\tau_0\right)+
  \frac{C_1}{2}\left[\wp\left(\tau-\tau_0\right)-\frac{C_2}{24}\right]+
  \frac{C_3}{24} f_0}{2\left[\wp\left(\tau-\tau_0\right)-\frac{C_2}{24}\right]^2-
  \frac{C_4}{48} f_0}\; ,  \label{sfmr5}
\end{equation}
where $\Psi(\tau_0)=\Psi_0$, $f_0=f(\Psi_0)$, $C_k=f^{(n)}(\Psi_0)$ is the $n$-th derivative of
$f$ at $\Psi=\Psi_0$ ($n=1,\ldots,4$), and ${\cal S}={\rm sgn}[\sin\psi_0]$ is the sign 
function ($1$ for $\sin\psi_0>0$ and $-1$ for $\sin\psi_0<0$). 
The time dependence of the solution is given in terms of the Weierstrass elliptic 
function $\wp(z)$, and $\wp^\prime(z)={\rm d}\wp(z)/{\rm d}z$, whose invariants are 
\begin{eqnarray}
 g_2 & = &  a_0a_4-4a_1a_3+3a_2^2 \; , \nonumber \\
 g_3 & = &  a_0a_2a_4+2a_1a_2a_3-a_2^3-a_0a_3^2-a_1^2a_4\; . \label{sfmr6b} 
\end{eqnarray}
Interestingly, as far as we know, the general solution (\ref{sfmr5}) 
has not been discussed in the literature. Ferraz-Mello (2007) mentioned a 
particular solution valid for $\sin\psi_0=0$, which implies that $f_0=0$ 
(see also Shinkin 1995). 

\begin{figure*}
 \epsscale{0.7}
  \plotone{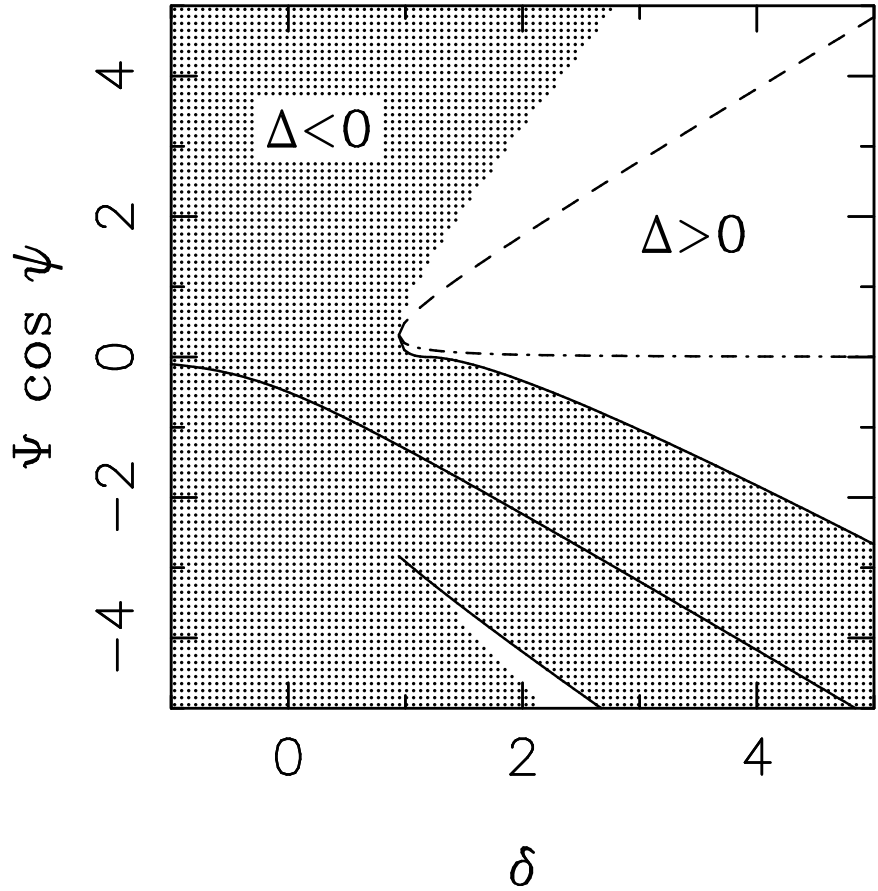}
 \caption{The range in parameter $\delta$ and $\Psi \cos \psi$ where
Eq. (\ref{sfmr5}) admits different functional dependence on the Jacobi elliptic functions.           
In the void domain, discriminant $\Delta>0$ and Eq. (\ref{sfmr7}) applies. In the dotted domain, 
$\Delta<0$ and Eq. (\ref{sfmr8}) applies. The lines show the location of the equilibrium 
points. See caption of Fig. \ref{delta2} more info.}
 \label{discrim}
\end{figure*}

The solution (\ref{sfmr5}) is most conveniently evaluated using the
relation of the Weierstrass functions to the Jacobi elliptic functions 
${\rm sn}$ and ${\rm cn}$. 
The form of the solution depends on the roots of the cubic equation 
$4z^3-g_2z-g_3=0$, whose discriminant is $\Delta=g_2^3-27 g_3^2$.
There are three real roots $e_1>e_2>e_3$ for $\Delta>0$. If $\Delta<0$, there is one 
real root $e_2$ and a pair of complex roots $e_1=\alpha+\imath \beta$ and 
$e_3=\alpha-\imath \beta$ (Figure \ref{discrim}). In the first case ($\Delta>0$), we 
have
\begin{equation}
 \wp\left(\tau-\tau_0\right) = e_3+\frac{e_1-e_3}{{\rm sn}^2\left(u,k\right)}
  \; ,  \label{sfmr7}
\end{equation}
with $u=\sqrt{e_1-e_3}\,(\tau-\tau_0)$ and modulus $k=\sqrt{\frac{e_2-e_3}{e_1-e_3}}$. 
In the second case ($\Delta<0$), we have
\begin{equation}
 \wp\left(\tau-\tau_0\right) = e_2+\gamma\,\frac{1+{\rm cn}\left(u,k\right)}{
  1-{\rm cn}\left(u,k\right)} \; ,  \label{sfmr8}
\end{equation}
with $u=2\sqrt{\gamma}\,(\tau-\tau_0)$, $\gamma=\sqrt{9\alpha^2+\beta^2}$ and
$k=\sqrt{\frac{1}{2}-\frac{3e_2}{4\gamma}}$. The Jacobi elliptic functions are computed 
following the numerical recipe from Press et~al. (2007). The derivative $\wp^\prime$ is
obtained from the derivatives of the Jacobi functions. 
\subsection{Resonant Period}
The solution (\ref{sfmr5}) is periodic with a period
\begin{equation}
 P_\tau = \frac{2\, \mathbb{K}\left(k\right)}{\sqrt{e_1-e_3}} \; ,  \label{sfmr9}
\end{equation}
for $\Delta>0$, and
\begin{equation}
 P_\tau = \frac{2\, \mathbb{K}\left(k\right)}{\sqrt{\gamma}} \; ,  \label{sfmr10}
\end{equation}
for $\Delta<0$. Here, $\mathbb{K}(k)$ is the complete elliptic integral of the first kind.

\begin{figure*}
 \epsscale{0.8}
  \plotone{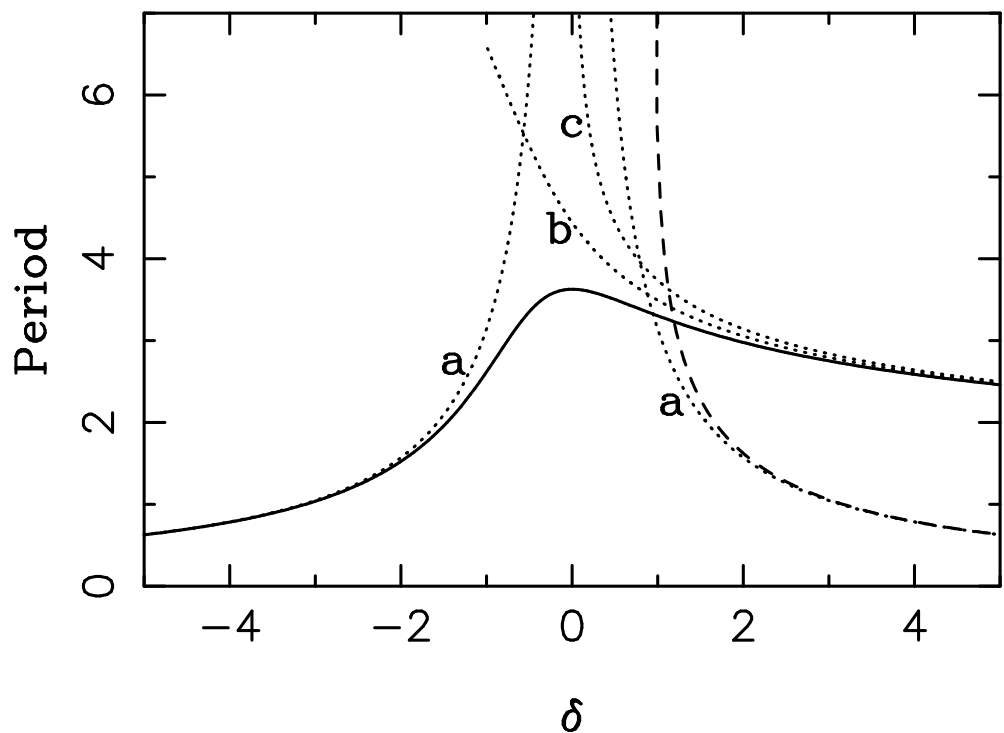}
 \caption{The period of small amplitude librations around the stable equilibrium points.
The solid and dashed lines show the periods in the libration (`stable 1' in 
Fig. \ref{delta2}) and circulation domains (`stable 2' in Fig. \ref{delta2}), respectively.
The dotted lines are various approximations. The dotted lines denoted by `a' 
is the super-period approximation with $P_\tau=\pi/|\delta|$. The one denoted by `b'
is an oscillator approximation $P_\tau = \pi (2/\Psi_{\rm eq})^{1/4}$, where 
$\Psi_{\rm eq}$ of the libration point is computed exactly for each $\delta$.
The dotted line `c' shows $P_\tau = \pi (2/\delta)^{1/4}$.}
 \label{period1}
\end{figure*}

Figure \ref{period1} shows the period of small amplitudes librations around the stable 
equilibrium points. Away from the resonance and if the eccentricities are small, 
$n_{\rm s}/(2 \nu) \gg \Phi_2$, and the first bracketed term 
in Eq. (\ref{delta}) outweighs the second. In this case, $\delta$ is related to the  
super frequency, and increases, in the absolute value, when the system moves away from 
the resonance. In this sense, $\delta$ is a measure of the 
distance from the resonance.\footnote{Inside the resonant island, $n_{\rm s}$ is small
and $\delta$ is controlled by the contribution from $\Phi_2$. Since $\Phi_2 \propto 
e^2$, $\delta$ is a measure of the orbital eccentricities inside the libration island.}

Negative values of $\delta$ imply that the planetary orbits are spaced more widely
than the actual resonance ($P_2/P_1 > k/(k-1)$), while $\delta>0$ means that the 
orbits are packed more tightly ($P_2/P_1 < k/(k-1)$). Also, $\delta \simeq 
\eta_2^{-1/3} n_{\rm s}/2$ away from the resonance, and therefore the scaled period 
$P_\tau \simeq \pi/|\delta|$. This represents a very good approximation of the period 
if $\delta<-2$ or $\delta>3$ (see Figure \ref{period1}). 

Inside the libration island for $\delta>\delta_*$, the period of small-amplitude 
librations decreases with $\delta$ (i.e., toward larger eccentricities). It can be 
approximated by $P_\tau = \pi (2/\Psi_{\rm eq})^{1/4}$, where $\Psi_{\rm eq}$ is the 
equilibrium value of $\Psi$ (Figure \ref{delta2}). For large values of $\delta$, 
$\Psi_{\rm eq}\simeq \delta$, 
and 
\begin{equation}
P_\tau \simeq \pi \left( { 2 \over \delta } \right)^{1/4} \; .
\end{equation}
Figure \ref{period1} shows that this approximation works very well for $\delta>3$. 

\begin{figure*}
 \epsscale{0.8}
  \plotone{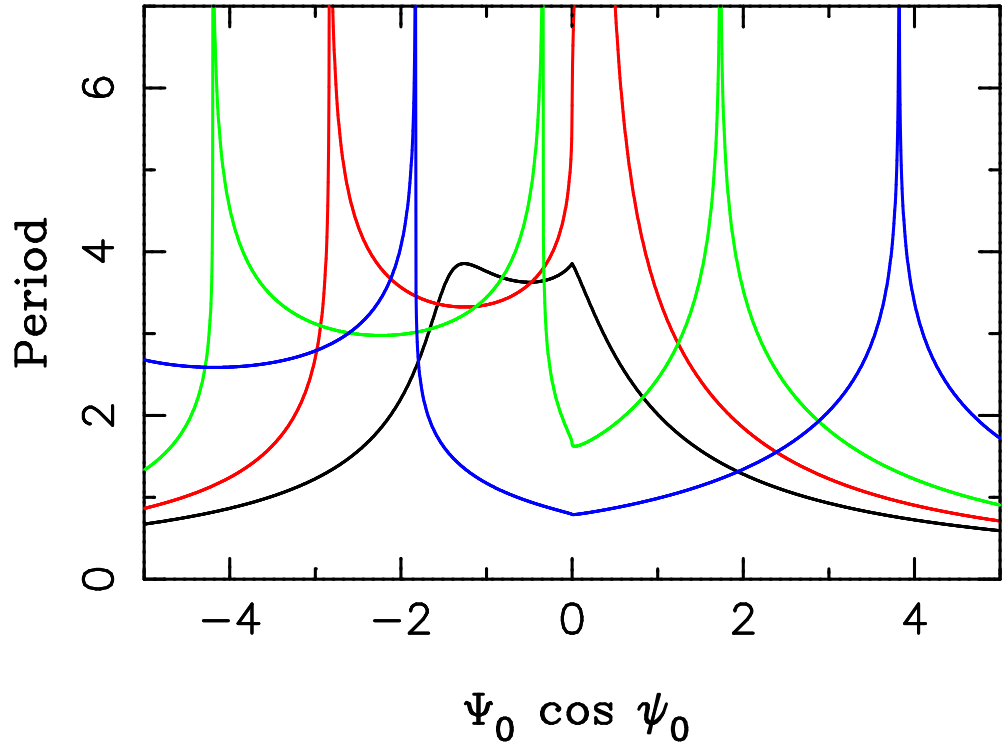}
 \caption{The period $P_\tau$ for different libration amplitudes. Here we set $\sin \psi_0=0$ 
and compute the period for different $\Psi_0$ from Eqs. (\ref{sfmr9}) and (\ref{sfmr10}). The different
colors correspond to different $\delta$ values: $\delta=0$ (black), $\delta=\delta_*$ (red),
$\delta=2$ (green), and $\delta=4$ (blue). This figure can be compared to Fig. \ref{portraits},
where the dynamical portraits are shown for the same values of~$\delta$.}
 \label{period2}
\end{figure*}

The period $P_\tau$ for different libration amplitudes is plotted in Fig. \ref{period2}.
The period is the shortest near the equilibrium point and increases with the libration 
amplitude. It becomes infinite on the separatrices. This corresponds to a situation 
when the amplitude of angle $\psi$ becomes full $\pi$, and it takes infinitely long
to reach the unstable equilibrium point at $\psi=0$. Except for trajectories near the
separatrix, the period values inside the libration island are $P_\tau\simeq2.5$-4.
They become shorter for $\delta>5$ (i.e., for higher eccentricities).  

When rescaled according to Eq. (\ref{tscale}), $P_t=\eta_2^{-1/3}P_\tau$ is the period
of resonant librations in the normal time units (e.g., Julian days). It has the same 
dependence of $P_\tau$ on $\delta$ and amplitude that we discussed above, and, in addition, 
contains an explicit dependence on the orbital period and planetary masses through the 
scaling parameter $\eta_2=\nu C^2$. If $m_1 \ll m_2$, then
\begin{equation}
P_t \simeq P_1\; {P_\tau \over 2 \pi} \left( {m_2 \over M_*} \right)^{\!\!-2/3} 
\left[ {3 \over 2} (k-1)^2 f_1^2 \alpha^2 \right]^{-1/3} \; . 
\label{perone}
\end{equation}
If, on the other hand, $m_1 \gg m_2$, then
\begin{equation}
P_t \simeq P_2\; {P_\tau \over 2 \pi} \left( {m_1 \over M_*} \right)^{\!\!-2/3} 
\left[ {3 \over 2} k^2 f_2^2 \right]^{-1/3} \; . 
\label{pertwo}
\end{equation} 
The multiplication coefficients in the square brackets are $\simeq$1 for $k=2$ (2:1 
resonance) and decrease with increasing $k$. This expresses a stronger interaction of 
orbits that are more tightly packed for larger values of $k$. According to Eqs. (\ref{perone}) 
and (\ref{pertwo}), the orbital periods $P_1$ and $P_2$ set the basic time unit for $P_t$.

\begin{figure*}
 \epsscale{0.47}
  \plotone{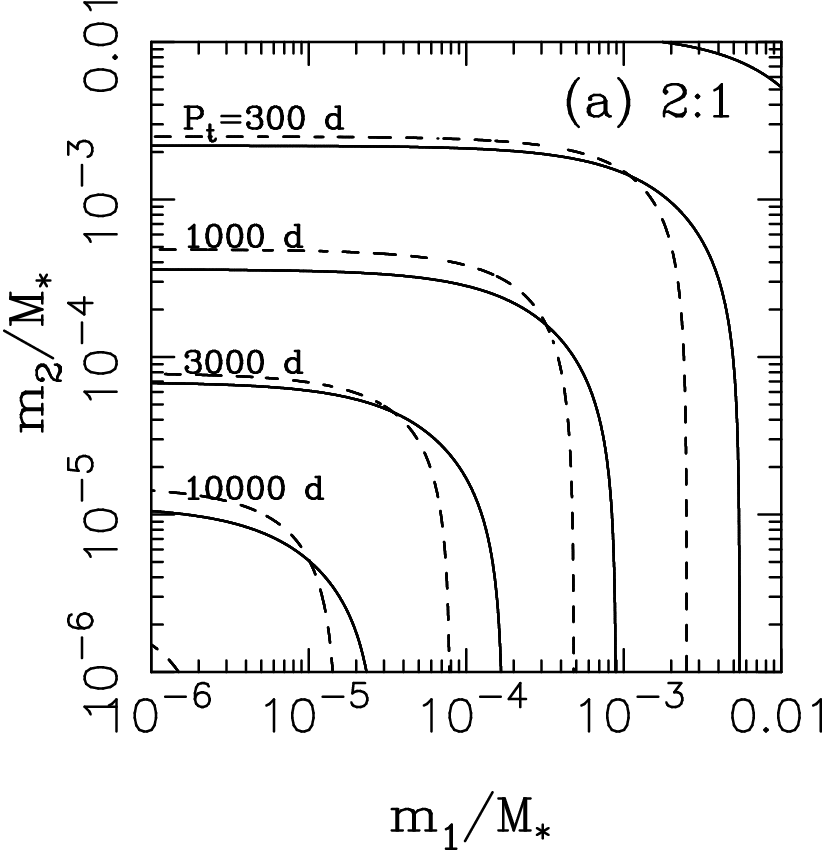}
  \plotone{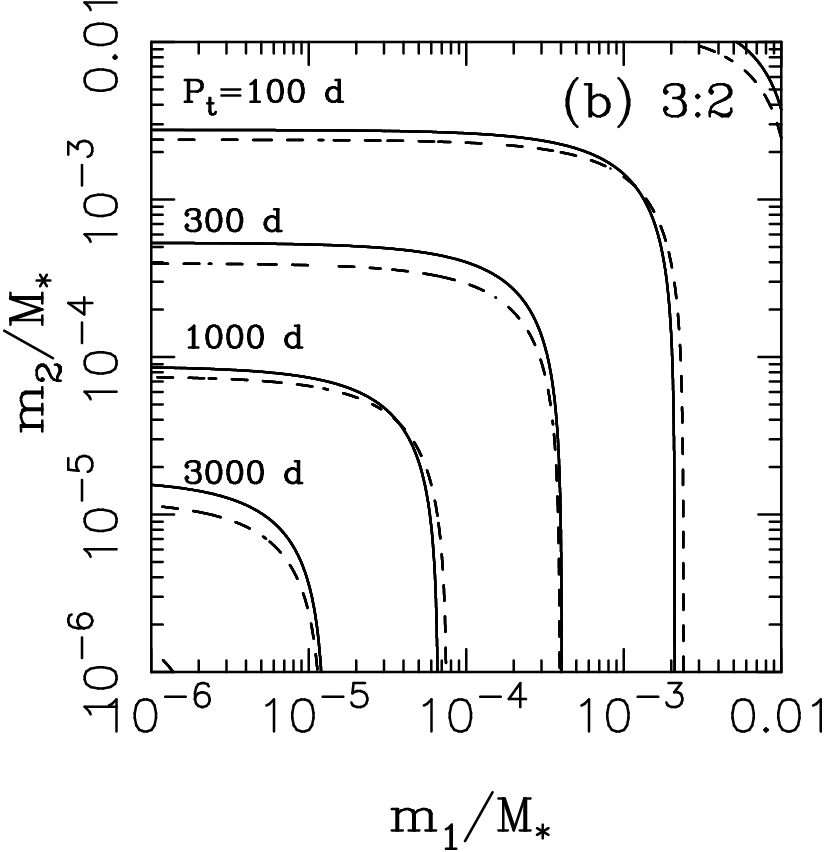}
 \caption{The libration period $P_t$ as a function of scaled planetary masses $m_1/M_*$ and $m_2/M_*$
(solid lines). Here we assumed that $P_1=10$ days and $P_\tau=3$ and computed $P_t=(\nu C^2)^{-1/3}P_\tau$.
Panels (a) and (b) show the results for the 2:1 and 3:2 resonances, respectively. This is 
the expected TTV period produced by the resonant librations in these resonances. The dashed lines 
show the approximation from Eq. (\ref{perthree}), which becomes better for larger values of $k$
(i.e., for $\alpha \rightarrow 1$).}
 \label{mass}
\end{figure*}

The libration period scales with the mass of the more massive planet as $(m/M_*)^{-2/3}$, 
and is therefore shorter for a larger mass. It is insensitive to the mass of 
the lighter planet. If the masses of the two planets are comparable, then $P_t$ will
depend on their combination via the scaling parameter $\eta_2$. Figure \ref{mass} 
shows the contour plots of $P_t$ as a function of $m_1/M_*$ and $m_2/M_*$. In the 
$\alpha\rightarrow1$ limit, the period is proportional to $[(m_1+m_2)/M_*]^{-2/3}$.
A reasonable approximation of the period is then
 \begin{equation}
P_t \simeq {P_1+P_2 \over 2} {P_\tau \over 2 \pi} \left( {m_1 + m_2 \over M_*} \right)^{\!\!-2/3} 
\left[ {3 \over 2} k(k-1) |f_1| f_2 \right]^{-1/3} \; . 
\label{perthree}
\end{equation} 
 
These considerations have important implications for the TTV period and 
its scaling with various parameters. For example, for two planets in the 2:1 
resonance with $m_2/M_* = 10^{-4}$ and $m_1 \ll m_2$, the expected TTV period is 
$P_t \sim 78 P_1 P_\tau$. Thus, if $P_1=10$~days, and assuming that $P_\tau\sim3$ in the 
libration regime, the TTV period will be $P_t\sim 6.4$ years. We can therefore very 
roughly write in this case 
\begin{equation}
P_t({\rm 2\!:\!1}) \sim 6.4\ {\rm yr} \times  {P_1 \over 10\ {\rm d} }  {P_\tau \over 3 } 
\left( {m_2  / M_*  \over 10^{-4} } \right)^{\!\!-2/3} \; .   
\end{equation} 
If, on the other hand, $m_1/M_* = 10^{-4}$ and $m_1 \gg m_2$, then the period will 
be approximately twice as long, because of having $P_2$ in Eq. (\ref{pertwo}) 
(instead of $P_1$ in Eq. (\ref{perone})). Finally, if $m_1/M_*=m_2/M_* = 10^{-4}$, then
$P_t({\rm 2\!:\!1})\sim4.5$ yr. The periods in the $k>2$ resonances are expected to 
be shorter (Figure \ref{mass}b).
\subsection{Approximation of Near-Resonant Orbits}
Lithwick et al. (2012; hereafter L12) derived elegant TTV expressions for two planets near (but not in) 
the first-order resonance. Here we discuss the relationship of these expressions to the 
results obtained here. Let us consider the Hamiltonian in Eq. (\ref{hk}) and (\ref{hper})  
\begin{equation}
{\cal H} = n_{\rm s} (\Gamma_1 + \Gamma_2) - \nu (\Gamma_1 + \Gamma_2)^2 + \beta_1 
\sqrt{2 \Gamma_1} \cos \sigma_1 + \beta_2  \sqrt{2 \Gamma_2} \cos \sigma_2 \, ,
\label{lit1}
\end{equation}
where we denote $\beta_1=-G m_1 m_2 A /a_2^*$ and $\beta_2=-G m_1 m_2 B /a_2^*$. We 
assume that the orbital eccentricities are small and that the two orbits are far enough 
from the resonance such that the term $\nu (\Gamma_1 + \Gamma_2)^2$ can be neglected. 
Introducing $x_j$ and $y_j$ from Eq. (\ref{ham10b}) into Eq. (\ref{lit1}) we find that 
\begin{equation}
{\cal H} = {n_{\rm s} \over 2} \left[ x_1^2+y_1^2+x_2^2+y_2^2\right] + \beta_1 x_1 + \beta_2 x_2\; .
\end{equation}
This Hamiltonian has a simple solution (e.g., Batygin \& Morbidelli 2013b). Defining complex 
variables $z_j= x_j + \imath y_j$, the solution can be written as
\begin{equation}
z_j=-{\beta_j \over n_{\rm s}} + C_j \exp (\imath n_{\rm s} t)\; ,
\label{lit2}
\end{equation}  
where $C_j$ are integration constants related to the initial conditions. Now, since
$\sigma_j=\theta-\varpi_j$, we can define L12's `complex eccentricities' $w_j=e_j \exp  
(\imath \varpi_j)$ and recast the approximate solution (\ref{lit2}) as
\begin{equation}
w_j = {\bar{C}_j \over \sqrt{\Lambda_1^*}} \exp \imath \theta_0 - {\beta_j \over n_{\rm s} \sqrt{\Lambda_1^*}} 
\exp \imath \theta\; ,
\label{lit3}
\end{equation} 
where $\bar{C}_j$ is the complex conjugate of $C_j$. Here we assumed that $\theta=\theta_0 + n_{\rm s}t$
and replaced $\Lambda_j \rightarrow \Lambda_j^*$. These assumptions are equivalent to those of   
L12 where the unperturbed (Keplerian) solution was inserted into the right-hand sides 
of the Lagrange equations, and the linearized solution was found by quadrature. 

Eq. (\ref{lit3}) can be compared to Eq. (A15) in L12. The first term in 
Eq. (\ref{lit3}) is constant and corresponds to the `free' term in Eq. (A15). The second term
in Eq. (\ref{lit3}) is identical to the second term in Eq. (A15) (this can be trivially 
shown by resolving notation differences). L12 proceeded by using the constants 
$K_1$ and $K_2$ (Eq. \ref{ham6b}) to derive expressions for $a_j(t)$, which were then used to 
obtain $\lambda_j= n_j t$ with $n_j=\sqrt{GM_*/a_j^3(t)}$.\footnote{The method of L12 ignores
the contribution to $\lambda_j$ from the derivatives of the Laplace coefficients. This is a 
correct assumption for the near-resonant orbits, because these terms have $n_{\rm s}$ in the 
denominator, while the terms from $n_j(t)$ have $n^2_{\rm s}$ in the denominator. The latter 
terms are thus amplified near a resonance where $n_{\rm s}$ is small.} Finally, L12 used 
a formula equivalent to Eq. (\ref{ttv}) to compute TTVs. None of these steps requires 
a special clarification.

The main assumption of L12 was therefore to (effectively) neglect the term $\nu (\Gamma_1 + \Gamma_2)^2$ 
in Eq.~(\ref{lit1}). For this to be valid $|n_{\rm s}| \gg \nu (\Gamma_1 + \Gamma_2)$. 
Assuming that the masses and eccentricities of the two planets are comparable, and neglecting all 
factors of the order of unity, this condition can be written as $e_j^2 \ll |P_2/P_1-k/(k-1)|$.
This shows that the eccentricities cannot exceed certain threshold for the L12 formulas to be valid.
This threshold is not excessively restrictive. For example, for planetary orbits just outside the 2:1 resonance
with $P_2/P_1-2=0.05$, the eccentricities need to be $e_j \ll \sqrt{0.05}=0.22$. 

\begin{figure*}
 \epsscale{0.50}
  \plotone{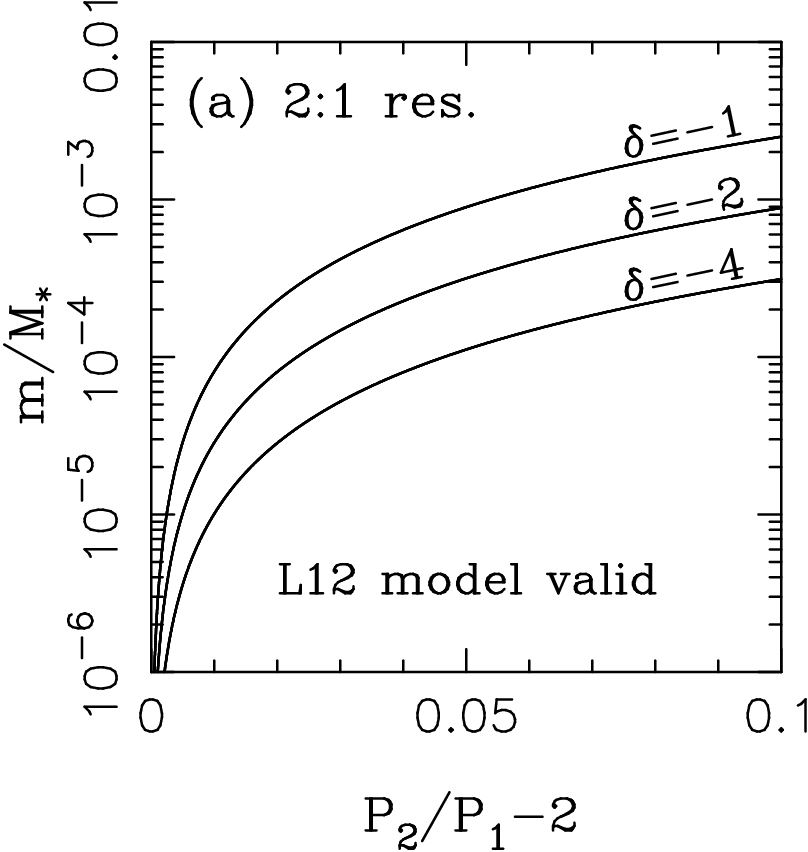}
 \epsscale{0.46}
  \plotone{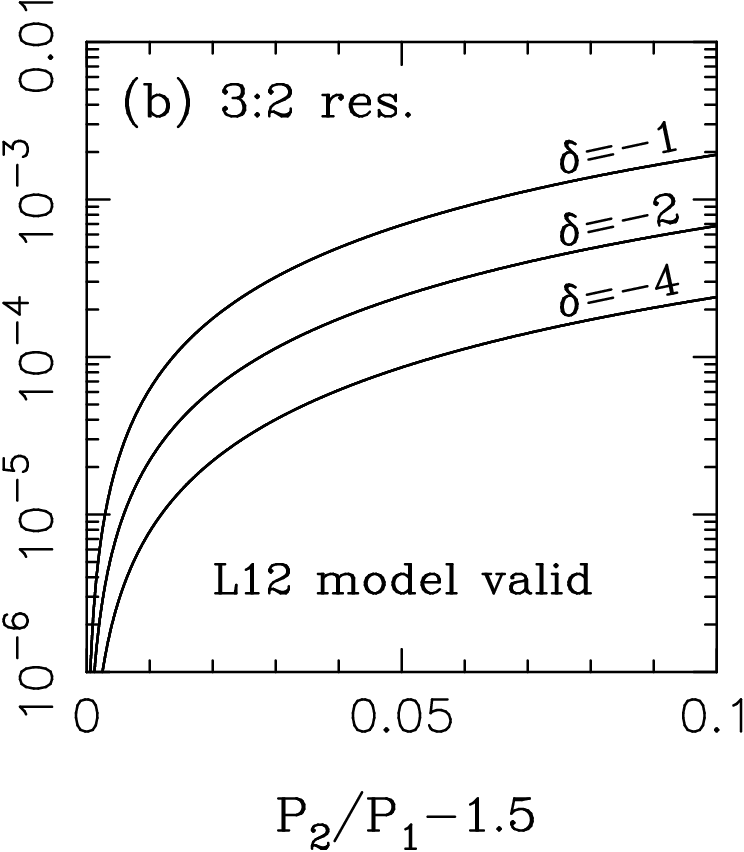}
 \caption{The validity domain of the L12 model. Here we assumed $m_1 \simeq m_2 = m$ 
and small orbital eccentricities, and plotted the isolines of $\delta$ from Eq. (\ref{delta}) as a 
function of planetary mass and $P_2/P_1-k/(k-1)$. According to the discussion in the main text,
the L12 model is valid for $\delta \lesssim -2$. This condition represents a combined constraint 
on the orbital period ratio and planetary masses. The parameter region below and to the right of 
the $\delta=-2$ line is where L12's TTV formula is strictly valid. This region covers most of
the period range shown here for planetary masses below that of Saturn.}
 \label{lit}
\end{figure*}

In addition to the above eccentricity condition, the planetary orbits cannot be too close to 
a resonance where the non-linear effects become important even for negligible eccentricities.
We tested this condition assuming initial orbits with small eccentricities and found that the L12
model is perfectly valid for $\delta<-4$. A small, $\sim$10\% discrepancy in both the amplitude 
and frequency of $w_j$ appears for $\delta\simeq -2$. Further increase of $\delta$ leads to a situation
where Eq. (\ref{lit3}) is no longer an adequate representation of the resonant dynamics. According 
to our tests with $e_1\simeq e_2 \lesssim 0.01$, the L12 model cannot usually be trusted for 
$\delta>-1$. Figure~\ref{lit} shows these thresholds as a function of the planetary mass and orbit 
period ratio.   
%
%
%
\section{Transit Timing Variations} 
If the variation of orbital elements is small, TTVs of two planets, $\delta t_1$ 
and $\delta t_2$, can be computed from
\begin{equation}
-n_j \delta t_j = \delta \lambda_j + 2 (\delta k_j \sin \lambda_j^* - \delta h_j 
\cos \lambda_j^*) + {\cal O}(e) 
\end{equation}
(see, for example, Nesvorn\'y \& Morbidelli 2008). Here, $k_j=e_j \cos \varpi_j$,
$h_j=e_j \sin \varpi_j$, and $\lambda_j=\lambda_j^*+\delta \lambda_j$, where 
$\lambda_j^* = n_j(t-t_0)$ with constant $n_j$. If the reference frame is chosen such that 
the orbital angles are measured with respect to the line of sight, transits occur 
when $\lambda_j^*\simeq0$ (assuming small eccentricities). We therefore have 
$-n_j \delta t_j = \delta \lambda_j - 2 \delta h_j + {\cal O}(e)$.
The first-order eccentricity terms, which can be used to improve the validity for higher 
eccentricities, were given in Nesvorn\'y (2009). 

TTVs will thus have a contribution from the mean longitude variation, 
$\delta \lambda_j$, and another contribution from the variation of eccentricity and
apsidal longitude, $\delta h_j=\delta(e_j \sin \varpi_j)$. Defining $Y_j=-\sqrt{2 \Gamma_j} 
\sin \varpi_j$, we have for small eccentricities that $\Gamma_j\simeq {1 \over 2} \Lambda_j e_j^2$
and $h_j=-Y_j/\sqrt{\Lambda_j}$. We then write $\Lambda_j=\Lambda_j^*+\delta \Lambda_j$ 
with constant $\Lambda_j^*$, and obtain
\begin{equation}
 -n_j \delta t_j = \delta \lambda_j + \frac{2}{\sqrt{\Lambda_j^*}} \delta Y_j + {\cal O}(e)\; ,
\label{ttvfin}
\end{equation}
where we retained only the first-order terms in small variations.

It remains to compute $\delta \lambda_j$ and $\delta Y_j$. As for $\delta \lambda_j$, we have
${{\rm d} \lambda_j / {\rm d} t} = {\partial {\cal H}_{\rm K} / \partial K_j}$ 
where ${\cal H}_{\rm K}$ is given in Eq. (\ref{hk}) (note that $K_j$ appears in the $n_{\rm s}$ 
term in (\ref{super})). Substituting $\Gamma_1+\Gamma_2 \rightarrow 
\Phi_1+\Phi_2$ in Eq. (\ref{hk}), taking the derivative with respect to $K_j$, and 
integrating with respect to $t$, we obtain
\begin{eqnarray}
\lambda_1 &=& \left[4 - {3 \over \Lambda_1^*}(K_1-(k-1)\Phi_2)\right]n_1(t-t_0)+3 (k-1) {n_1 \over 
\Lambda_1^*} \int_{t_0}^t \Phi_1 {\rm d}t\; , \nonumber \\
\lambda_2 &=& \left[4 - {3 \over \Lambda_2^*}(K_2+k\Phi_2)\right]n_2 (t-t_0)-3 k {n_2 \over 
\Lambda_2^*} \int_{t_0}^t \Phi_1 {\rm d}t\; .
\label{lamint}
\end{eqnarray}
Using the scaling relationship from Eqs. (\ref{psi}) and (\ref{tscale}), we have that
\begin{equation}
\int_{t_0}^t \Phi_1(t) {\rm d}t = {1 \over \nu} \int_{\tau_0}^\tau \Psi(\tau) {\rm d}\tau\; ,
\label{iscale}
\end{equation}
where $\Psi(\tau)$ is given in Eq. (\ref{sfmr5}). The first terms in (\ref{lamint}) describe
a uniform circulation of angles $\lambda_1$ and $\lambda_2$ (note that $4 n_j-3 K_j (n_j 
/\Lambda_j^*) = n_j$ for $e_j=0$, as expected). They will not contribute to TTVs. Instead, TTVs will 
arise from the integral term. As the sign in front of the integral term is positive for 
$\lambda_1$ and negative for $\lambda_2$, TTVs of the two planets are anti-correlated. 
The amplitudes of $\delta t_1$ and $\delta t_2$, denoted here by $A_{\lambda,1}$ 
and $A_{\lambda,2}$, satisfy
\begin{equation}
{A_{\lambda,1} \over A_{\lambda,2}} = { k-1 \over k } {\Lambda_2^* \over \Lambda_1^*}
\simeq \left({k-1 \over k}\right)^{2/3} {m_2 \over m_1}\; .  
\label{ratio1}
\end{equation}
Thus, apart from a coefficient of the order of unity, the ratio of the TTV amplitudes from
the $\lambda$ terms is expected to be equal to the inverse of the planetary mass ratio. 

As for $\delta Y_j$, we have
\begin{eqnarray}
Y_1 & = & { 1 \over \sqrt{A^2 + B^2} } 
\left[A (v_1 \cos \theta - u_1 \sin \theta) + B V_2\right]\; , \nonumber \\
Y_2 & = & { 1 \over \sqrt{A^2 + B^2} } 
\left[B (v_1 \cos \theta - u_1 \sin \theta) - A V_2\right]\; ,
\label{ys}
\end{eqnarray}  
where $u_1$ and $v_1$ were defined in Eq. (\ref{redt}). They are related to the $\Phi_1$ and $\phi_1$
variables via Eq. (\ref{ham14b}). Expressions (\ref{ys}) are derived in Appendix B. 
As we show in Appendix B, $V_2$ is a constant of motion, does not contribute to variations, 
and does not need to be computed. 

Since the coefficients $f_1$ and $f_2$ have opposite signs (Table 1), $A$ and $B$ 
in Eq. (\ref{ys}) will also have opposite signs, and TTVs of the two planets will be anti-correlated. 
TTVs arising from terms in (\ref{ys}) will have amplitudes, $A_{h,1}$ and $A_{h,2}$, such that
\begin{equation}
{A_{h,1} \over A_{h,2}} = {n_2 \over n_1} {A \over B} \sqrt{\Lambda_2^* \over \Lambda_1^*}
\simeq \left({k-1 \over k}\right)^{2/3} {f_1 \over f_2} {m_2 \over m_1}\; .  
\end{equation}  
The ratio of TTV amplitudes from the $h$ terms is inversely proportional to 
the ratio of planetary masses. This is the same mass dependence as in Eq. (\ref{ratio1}). 
Thus, while resonant TTV period discussed in Section 3.7 can be used 
to constrain $m_1/M_*$ and/or $m_2/M_*$, the TTV amplitude ratio is sensitive to $m_2/m_1$. 
     
Above we reduced the problem in hand to the evaluation of $v_1 \cos \theta - u_1 \sin \theta$.
The $\theta$ terms are simple. From Eq. (\ref{lamint}) we have that
\begin{equation}
\theta = [n_{\rm s} - 2 \nu \Phi_2](t-t_0) - 2 \nu \int_{t_0}^t \Phi_1 {\rm d}t 
\label{theta}
\end{equation}
As for the $u_1$ and $v_1$ terms, we obtain
\begin{eqnarray}
u_1 =  \eta_1^{-1/3} \sqrt{2\Psi}\,\cos\psi & = &  
-\eta_1^{-1/3}\left[h_0+\left(\Psi-\delta\right)^2
  \right]\; , \label{sfmr11a} \nonumber \\
v_1 =  \eta_1^{-1/3} \sqrt{2\Psi}\,\sin\psi & = &  
-\eta_1^{-1/3}\frac{d\Psi}{d\tau}\; . \label{sfmr11b} 
\label{uv1}
\end{eqnarray}

This is all we need for the computation of TTVs. To summarize, $\Psi(\tau)$ from Eq.~(\ref{sfmr5}) needs
to be inserted in (\ref{iscale}) and (\ref{uv1}). The integral in (\ref{iscale}) needs 
to be computed and substituted into Eqs. (\ref{lamint}) and (\ref{theta}). Then, $\theta$, 
$u_1$ and $v_1$ obtained from Eqs. (\ref{theta}) and (\ref{uv1}) are substituted into
(\ref{ys}). The constant terms in $\lambda_j$ and $Y_j$ can be neglected, because they
do not contribute to TTVs. Finally, Eq. (\ref{ttvfin}) is used to compute TTVs.  

In principle, the method outlined above can be used to derive fully analytic 
expressions for $\delta t_1$ and $\delta t_2$, which would have the same general 
validity for low eccentricities as the analytic solution (\ref{sfmr5}). For that, 
however, we would need to compute the derivative (needed for $v_1$) and integral 
(needed for $\lambda_j$ and $\theta$) of $\Psi(\tau)$ from (\ref{sfmr5}). The integral 
ends up producing complex expressions (Appendix C). Here we therefore opt for a different 
approach, where we seek to find an expression for resonant TTVs in terms of the
Fourier series.
\subsection{Fourier Series Expansion}
\begin{figure*}
 \epsscale{0.8}
  \plotone{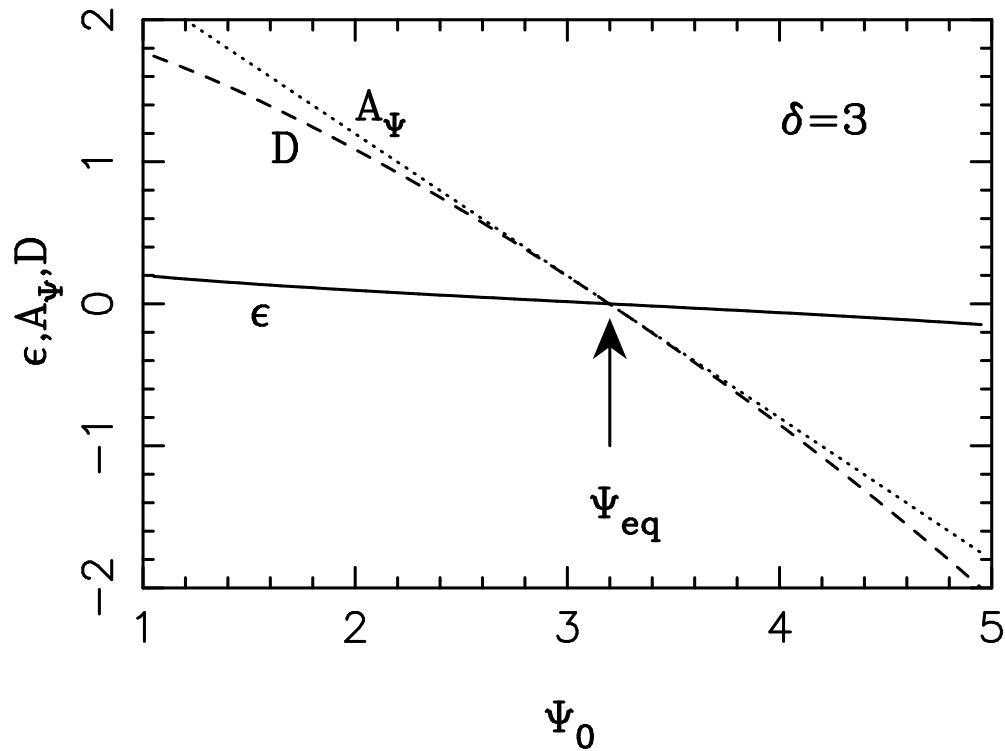}
 \caption{The dependence of various parameters of the analytic model on initial conditions. 
Here we set $\delta=3$, $\psi_0=\pi$, and vary $\Psi_0$. The solid line shows the small
parameter $\epsilon=b/a$. It is zero at the equilibrium point $\Psi_0=\Psi_{\rm eq}$ 
(labeled by the arrow) and increases to $\epsilon \simeq 0.2$ near the separatrices. 
The dotted line shows the amplitude $A_\Psi=\Psi_0-\Psi_{\rm eq}$. The parameter $D=C_1/4 a$
(Section 4.1), showed by the dashed line, is an excellent approximation of $A_\Psi$ for small 
amplitude librations.}
 \label{eps}
\end{figure*}

The solution (\ref{sfmr5}) is valid for any initial condition $(\psi_0,\Psi_0)$. Here we 
are not primarily interested in finding a general TTV expression for any $(\psi_0,\Psi_0)$. 
Instead, our primary goal is to understand the general scaling of TTVs with planetary masses, 
resonant amplitude, etc. We therefore opt for setting $\sin \psi_0 = 0$. 
This simplifies (\ref{sfmr5}) considerably. Specifically, $\sin \psi_0 = 0$ implies 
that $f_0=0$, and (\ref{sfmr5}) becomes
\begin{equation}
\Psi(\tau)=\Psi_0 + {C_1 \over 4} { 1 - {\rm cn}(u,k) \over a + b\; {\rm cn}(u,k) }\; ,
\label{simple}
\end{equation}
where we denoted $a = \gamma + e_2 - C_2/24$, $b = \gamma  - e_2 + C_2/24$, and 
$\gamma=\sqrt{9 \alpha^2 + \beta^2}$. This equation is valid in the domain $\Delta<0$ shown in
Fig. \ref{discrim}, which includes the whole resonant libration zone. We used Eq.
(\ref{sfmr8}) to relate the Weierstrass functions in (\ref{sfmr5}) to the Jacobi functions.  
 At the equilibrium point, $\Delta=0$, $b=0$, and 
$a = 2(\Psi_{\rm eq}-\delta)^2+\sqrt{2 \Psi_{\rm eq}}$ (Appendix A). For 
librations around the equilibrium point, we have that $a \gg b$ (Figure \ref{eps}). 
We therefore identify a small parameter $\epsilon = b/a \ll 1$ and expand 
Eq.~(\ref{simple}) in the Taylor series in $\epsilon$. 

Retaining only the first-order terms in $\epsilon$ (approximation of low-amplitude
librations), we obtain
\begin{equation}
\Psi(\tau)=\Psi_0 + {C_1 \over 4 a} \left[1-(1+\epsilon){\rm cn}(u,k) + \epsilon\; {\rm cn}^2(u,k)
\right] + {\cal O}(\epsilon^2)\; .
\label{approx1}
\end{equation}
Including higher order terms in $\epsilon$ would improve the validity of the 
approximation for large libration amplitudes. Next, we express ${\rm cn}$ and ${\rm cn}^2$ in 
the Fourier series
\begin{equation}
{\rm cn}(u,k)={2 \pi \over k \mathbb{K}} \sum_{n=1}^\infty {q^{n-1/2} \over 1 + q^{2n-1}}
\cos { (2n-1)\pi \over 2 \mathbb{K} } u 
\label{four1}
\end{equation}
and 
\begin{equation}
{\rm cn}^2(u,k)= {\mathbb{E} - k'^2 \mathbb{K} \over k^2  \mathbb{K} } +   
{2 \pi^2 \over k^2 \mathbb{K}^2} \sum_{n=1}^\infty {n q^{n} \over 1 - q^{2n}}
\cos { n \pi \over \mathbb{K} } u \; .
\label{four2}
\end{equation}
Here we have $k'=\sqrt{1-k^2}$ and $q=\exp(-\pi \mathbb{K}'/\mathbb{K})$ with 
$\mathbb{K}'=\mathbb{K}(k')$. $\mathbb{E}(k)$ is the complete elliptic integral of the second kind.
General expressions for ${\rm cn}^m(u,k)$, which can 
become useful when higher order terms in (\ref{approx1}) are accounted for, can be found in 
Kiper (1984). 

Both these Fourier series converge very rapidly, and we can therefore afford to 
use the lowest harmonics. In practice, given that the evaluation of $u_1$ from Eq. 
(\ref{uv1}) will require a multiplication of the Fourier series, here we consider only 
the first and second harmonics of $u=2\sqrt{\gamma}(\tau-\tau_0)$. After substituting 
these terms into Eq. (\ref{approx1}), we obtain 
\begin{equation}
\Psi=\Psi_0 + D \left (1+ {\epsilon \over 2}\right) - D(1+\epsilon) \cos f_\tau \tau + 
{1 \over 2} D \epsilon \cos 2 f_\tau \tau\; .
\label{psi1}
\end{equation}
Here we denoted the frequency $f_\tau = 2 \pi / P_\tau = \pi \sqrt{\gamma} / \mathbb{K}$
and $D=C_1/4a$. To simplify things, we set $\tau_0=0$ and drop all multiplication
terms appearing from Eqs. (\ref{four1}) and (\ref{four2}) that are $\simeq 1$.\footnote{
Specifically, except for $k$ very close to 1, we have $(\mathbb{E}-k'^2 \mathbb{K})/(k^2 \mathbb{K}) 
\simeq 1/2$, $2 \pi \sqrt{q} / [k\mathbb{K} (1+q)] \simeq 1$, and 
$2 \pi^2 q/ [k^2\mathbb{K}^2 (1-q^2)] \simeq 1/2$.}  
 
Note that $D$ is a proxy for the libration amplitude of $\Psi$ (Figure \ref{eps}).
Eq.~(\ref{psi1}) can be used to trivially compute the integral $\int \Psi 
{\rm d}\tau$ appearing in Eq. (\ref{iscale}), which is then inserted into Eqs. 
(\ref{lamint}) and (\ref{theta}). In the following text, the two TTV contributions, 
$\delta \lambda_j$ and $\delta Y_j$, will be considered separately. 
\subsection{Contribution from Mean Longitude Variations}  
The calculation of $\delta \lambda_j$ is simple. Retaining the periodic terms in Eq.
(\ref{lamint}) we obtain
\begin{eqnarray}
\delta \lambda_1 & = & - 3 (k-1) { n_1 \over \Lambda_1^* \nu }  {P_\tau \over 2 \pi} D
[ C_{\lambda,1} \sin f t + C_{\lambda,2} \sin 2 f t ]\; , \nonumber \\
\delta \lambda_2 & = & 3 k { n_2 \over \Lambda_2^* \nu }  {P_\tau \over 2 \pi} D
[ C_{\lambda,1} \sin f t + C_{\lambda,2} \sin 2 f t ]\; .
\label{dl2}
\end{eqnarray}
with the coefficients $C_{\lambda,1}  =  1+\epsilon$ and $C_{\lambda,2}  =  - \epsilon/4$, and
frequency $f=\eta_2^{1/3} f_\tau= \eta_2^{1/3} \pi \sqrt{\gamma} / \mathbb{K}$ (Section 3.7).

These equations are the source of Eq. (\ref{dl}) in Section 2, where we give the final expressions 
for TTVs arising from the variation of $\lambda_1$ and $\lambda_2$. In Eq. (\ref{dl}),
we have taken the liberty to drop the star from $\Lambda_j^*$, but it is understood that
$n_j$, $\Lambda_j$, and other quantities depending on these parameters in Eq. (\ref{dl}) are considered 
to be constant. Also, since $D$ is a good proxy for the amplitude $A_\Psi$, we replaced 
$D \rightarrow A_\Psi$ in Eq. (\ref{dl}). Note that $A_\Psi$ is positive for 
$\Psi_0>\Psi_{\rm eq}$ and negative for $\Psi_0<\Psi_{\rm eq}$. 
\subsection{Contribution from Eccentricity and Apsidal Longitude}
We need to compute $v_1 \cos \theta - u_1 \sin \theta$ and insert it in Eq. (\ref{ys}).
As for the terms including $\theta$, we obtain from Eqs. (\ref{iscale}), (\ref{theta}) and  (\ref{psi1}) 
\begin{equation}
\theta=\theta_0 + f_\theta t + C_{\theta,1} \sin f t + C_{\theta,2} \sin 2 f t\; , 
\end{equation}  
where
\begin{eqnarray}
f_\theta & = & n_{\rm s} - 2 \nu \Phi_2 -2 \eta_2^{1/3} \left[\Psi_0+
D\left(1+{\epsilon \over 2}\right)\right]\; , \nonumber \\
C_{\theta,1} & = & {P_\tau \over 2 \pi} D (1+\epsilon) \; , \nonumber \\
C_{\theta,2} & = & - { P_\tau \over 2 \pi} D \epsilon\; .
\label{ftheta}
\end{eqnarray}
We then use the following expansions to obtain $\cos \theta$ and $\sin \theta$
\begin{eqnarray}
\cos(x\sin\varphi) & = &J_0(x) + 2 \sum_{n=1}^\infty J_{2n}(x) \cos 2 n \varphi\; , \nonumber \\
\sin(x\sin\varphi) & = &2 \sum_{n=1}^\infty J_{2n-1}(x) \sin (2 n-1)\varphi\; , 
\end{eqnarray}
where $J_n(x)$ are the Bessel functions. Both these Fourier series converge rapidly
for $x \ll 1$. We therefore retain only the lowest-order harmonics.
This leads to
\begin{eqnarray}
\cos \theta & = & J_0(C_{\theta,1}) \cos(\theta_0+f_\theta t)\nonumber \\ 
& - & 2 J_1(C_{\theta,1}) \sin(\theta_0+f_\theta t) \sin ft +
2 J_2(C_{\theta,1}) \cos(\theta_0+f_\theta t) \cos 2ft\; , \nonumber \\
 \sin \theta & = & J_0(C_{\theta,1}) \sin(\theta_0+f_\theta t)\nonumber \\ 
& + & 2 J_1(C_{\theta,1}) \cos(\theta_0+f_\theta t) \sin ft
+ 2 J_2(C_{\theta,1}) \sin(\theta_0+f_\theta t) \cos 2ft\; . 
\label{cossin}
\end{eqnarray}
Here we neglected the coefficients $C_{\theta,2}$ that are of the order of $\epsilon$ 
(note that $\epsilon \ll D$; Fig. \ref{eps}).  

The harmonics with $f$ in Eq. (\ref{cossin}) appear from the resonant librations (Section 3.7). 
Since $\theta=\phi_1-\zeta_1$ (Appendix B), and $\phi_1$ oscillates around $\pi$ in the 
libration island, $\theta$ has the same circulation frequency as $\zeta_1$. Now, given 
that $\zeta_1$ is defined from the apsidal longitudes $\varpi_1$ and $\varpi_2$ (Appendix B),
the interpretation of $f_\theta$ is that it is the mean precession frequency of the 
longitudes of periapsis. Neglecting terms ${\cal O}(\epsilon)$ and using the definition 
of $\delta$ in Eq.~(\ref{delta}) we find from (\ref{ftheta}) that 
\begin{equation} 
f_\theta=2 \eta_2^{1/3} (\delta - \Psi_{\rm eq})\; ,
\label{ftheta2}
\end{equation}
where substituted $\Psi_0+D \rightarrow \Psi_{\rm eq}$ (see Figure \ref{eps}). The 
$f_\theta$ frequency therefore scales in the same manner with planetary parameters 
as $f=\eta_2^{1/3} f_\tau$ (as expressed by the $\eta_2^{1/3}$ factor). Unlike $f$,
which derives from $f_\tau=2\pi/P_\tau\sim 3$ in the libration island, $f_\theta$ contains 
the factor $2 (\delta - \Psi_{\rm eq})$. This factor is $\ll 1$ (compare the 
dotted and solid `stable 1' lines in Figure \ref{delta2}). Therefore, 
$f_\theta$ is substantially smaller than $f$, which shows that the variations from $\theta$ 
are expected to occur on a long timescale.

Specifically, the period $P_\theta=2 \pi/f_\theta =\pi / (\delta - \Psi_{\rm eq})$ is equal to 10.1 
for $\delta=1$, 13.3 for $\delta=2$, and 18.1 for $\delta=4$. The longer periods 
for larger $\delta$ values are a consequence of $\Psi_{\rm eq}$ approaching $\delta$ for 
increasing values of $\delta$ (Section 3.5). Note that $P_\theta$, at 
least in the approximation adopted here, is independent of the libration amplitude. 
Also, given that $\delta - \Psi_{\rm eq}<0$ in the libration island, the $f_\theta$ 
frequency is negative as well, meaning that the circulation of $\theta$ is retrograde 
(implying retrograde rotation of $\varpi_1$ and $\varpi_2$). 

The terms in Eq. (\ref{cossin}) containing frequencies $f_\theta$ and $f$ could be combined 
together to produce harmonics with frequencies $f_\theta \pm f$ and $f_\theta \pm 2 f$. Given that, 
as we discussed above, the characteristic periods of these terms are largely different, we
prefer to leave them multiplying each other in Eq. (\ref{cossin}). Accordingly, Eq. 
(\ref{cossin}) is interpreted as the resonant variations around the mean value that is 
slowly modulated with frequency $f_\theta$. 

\begin{figure*}
 \epsscale{0.7}
  \plotone{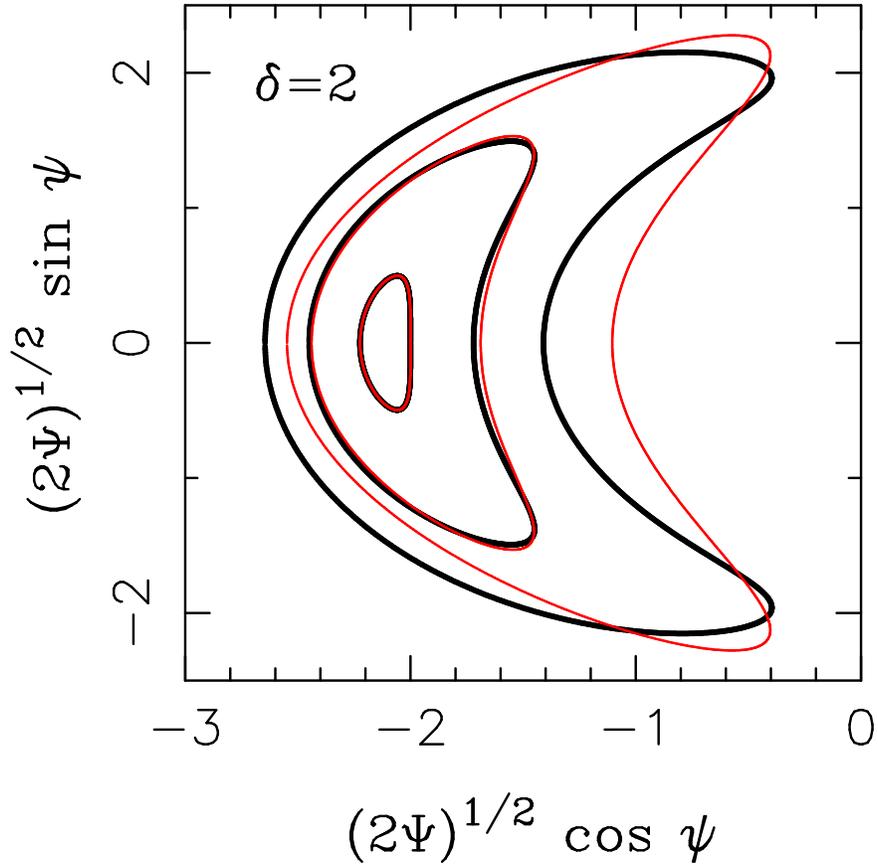}
\caption{This plot illustrates the approximation (\ref{uvapp}). The bold black lines are 
the exact solution of the resonant Hamiltonian for $\delta=2$ and three libration 
amplitudes. The red lines are the approximation given in Eq. (\ref{uvapp}). The approximation 
is good for small libration amplitudes (inner curves) and degrades when the 
amplitude increases (outer curve).}
\label{uvfig}
\end{figure*}

The expressions for $u_1$ and $v_1$ are derived from Eq.~(\ref{uv1}), after substituting $\Psi$
from Eq.~(\ref{psi1}). After some algebra we obtain
\begin{eqnarray}
u_1 & = & \eta_1^{-1/3} D \left(C_{u,0} + C_{u,1} \cos f t + C_{u,2} \cos 2 f t 
+ C_{u,3} \cos 3 f t\right)\; ,  \nonumber \\
v_1 & = & \eta_1^{-1/3} {2 \pi \over P_\tau } D \left(-(1+\epsilon) \sin f t + \epsilon \sin 2 f t\right)
\label{uvapp}
\end{eqnarray}
with coefficients
\begin{eqnarray}
C_{u,0} & = & {u_{0} \over D} - D \left( {3 \over 2} + 2 \epsilon \right) - (2+\epsilon)(\Psi_0-\delta)\; ,\nonumber \\
C_{u,1} & = & D\left(2 + {7 \over 2} \epsilon\right) + 2(1+\epsilon)(\Psi_0-\delta)\; , \nonumber \\
C_{u,2} & = & -D\left( {1 \over 2} + 2 \epsilon \right) -\epsilon( \Psi_0 - \delta)\; , \nonumber \\
C_{u,3} & = & {1 \over 2} \epsilon D\; .
\label{cus}
\end{eqnarray}
In the above expression for the coefficient $C_{u,0}$, $u_0$ denotes the initial value 
\begin{equation}
u_{0}=\sqrt{2 \Psi_0}=-h_0-(\Psi_0-\delta)^2\; .
\end{equation}
Figure \ref{uvfig} compares Eq.~(\ref{uvapp}) with the exact solution of Eqs. (\ref{hameq1})
and (\ref{hameq2}). 
It shows that the approximation (\ref{uvapp}) is excellent for small libration amplitudes but
loses precision for large libration amplitudes. This happens mainly because the terms ${\cal O}(\epsilon^2)$ 
were neglected in Eq.~(\ref{approx1}). In principle, it should be possible to include
these and higher order terms and improve the validity of the Fourier approximation by producing
more general expressions. We leave this for future work. 

Now we should combine Eqs. (\ref{cossin}) and (\ref{uvapp}) together. Unfortunately, this generates 
a very long expression for $\delta Y_j$. We do not explicitly give this equation here. The full expression 
was coded in a program and used to generate Figs. \ref{ttv1} and \ref{ttv2}. 
We find that the TTV terms from $v_1 \cos \theta - u_1 \sin \theta$ with the frequency $f$ have amplitudes 
that are generally much smaller than the TTV amplitudes arising from the $\delta \lambda_j$ terms 
(Eq. \ref{dl2}). Here we therefore explicitly report only the most important harmonic with frequency
$f_\theta$. These terms do not have a counterpart in Eq. (\ref{dl2}). They are important for the long-term
modulation of the TTV signal. Specifically, we find that 
\begin{eqnarray}
\delta Y_1 & = & -\eta_1^{-1/3} {A D \over \sqrt{A^2 + B^2}} C_{u,0} J_0(C_{\theta,1}) 
\sin(\theta_0+f_\theta t)\; , \nonumber \\
\delta Y_2 & = & -\eta_1^{-1/3} {B D \over \sqrt{A^2 + B^2}} C_{u,0} J_0(C_{\theta,1}) 
\sin(\theta_0+f_\theta t)\; .
\label{yys}
\end{eqnarray}
This is the source of Eq. (\ref{yys2}). Since $J_0(C_{\theta,1})\sim1$,
we do not list this term in Eq. (\ref{yys2}), where we also substitute $D \rightarrow A_\Psi$. 
\section{The Domain of Validity} 
We adopted several approximations in this work:
\begin{description}
\item{\bf I.} The Laplacian expansion of the perturbing function used in Section 3.1 is convergent 
only if the planetary eccentricities are small enough (Sundman 1912). For a planet on a circular 
orbit, this limits the validity of the expansion to $e<0.25$ ($e<0.2$) for orbits near its 
inner (outer) 2:1 resonance, and to $e<0.15$ ($e<0.12$) for orbits near its inner (outer) 
3:2 resonance. The analytic results derived here are not valid above these 
limits. See Nesvorn\'y \& Morbidelli (2008) for a discussion of Sundman's criterion.
\item{\bf II.} All non-resonant terms were neglected in Section 3.1. The short-periodic terms with non-resonant 
frequencies produce short-periodic TTVs that can be calculated by the method described in Nesvorn\'y \& Morbidelli 
(2008). These terms can be linearly added to the expressions obtained here for resonant TTVs. The secular
terms are second and higher orders in planetary eccentricities and contribute by only a small 
correction to the precession of orbits if the eccentricities are small. 
\item{\bf III.} The second- and higher-order resonant terms in planetary eccentricities were neglected in 
the perturbing function. This is an important approximation that limits the validity of the results
to small eccentricities. The same assumption was adopted when writing $\delta t_j$ as a variation 
of orbital elements. 
\item{\bf IV.} The amplitude of the semimajor axis variations was assumed to be small (this allowed us
to simplify the Keplerian Hamiltonian in Section 3.3). The same assumption was adopted to compute
$\delta t_j$ from Eq. (\ref{ttv}), where we also neglected all second and higher order terms in small variations of the orbital 
elements. We find that these approximations are generally valid and do not impose any meaningful limits 
on the range of planetary parameters where our analytic results are valid.  
\item{\bf V.} The exact solution of the second fundamental model of resonance was expanded in the Taylor series 
in $\epsilon$ and only the terms ${\cal O}(\epsilon)$ were retained (Section 4.1). In addition, the Jacobi elliptic 
functions were written as the Fourier series and only the lowest harmonics were retained. Both these
approximations limit the validity of our analytic TTV model to relatively small libration amplitudes. 
[In principle, the methods described in Section 4 can be used to obtain more general expressions.]
\end{description}

Here we perform tests of these assumptions to establish the domain of validity of our analytic model.
To this end we developed several codes that compute the resonant TTV signal at various stages of approximation. 
They are:
\begin{description}
\item{\bf A.} A full $N$-body integrator of Eq. (\ref{ham1}) and (\ref{ham2}) where the gravitational interaction 
of planets is taken into account exactly. We used the symplectic code known as {\tt Swift} (Levison \& Duncan 
1994) with routines for an efficient and precise determination of TTVs (Nesvorn\'y et al. 2013; see 
also Deck et al. 2014).
\item{\bf B.} A numerical integrator in orbital elements that uses the Laplacian expansion of the perturbing 
function. Various terms can be included or excluded in this integrator. In the most basic approximation, the 
code includes only the first-order resonant terms from Eq. (\ref{ham4}). Optionally, it also accounts for the 
second-order secular and/or resonant terms. This code is used to test the approximations {\bf II} and {\bf III} 
listed above  
\item{\bf C.} A code that maps the initial orbital elements onto (\ref{fm}) and numerically integrates the 
corresponding Eqs. (\ref{hameq1}) and (\ref{hameq2}). Another code uses the exact analytic solution 
(\ref{sfmr5}). As expected, these two codes give exactly the same result, which shows that our
implementation of Eq. (\ref{sfmr5}) is working correctly.
\item{\bf D.} A TTV code based on the analytic formulas derived in Section 4. This code is subject to all 
approximations discussed above. It cannot produce accurate results if the orbital 
eccentricities and/or libration amplitudes exceed certain limits.
\end{description}   

\begin{figure*}
 \epsscale{0.85}
  \plotone{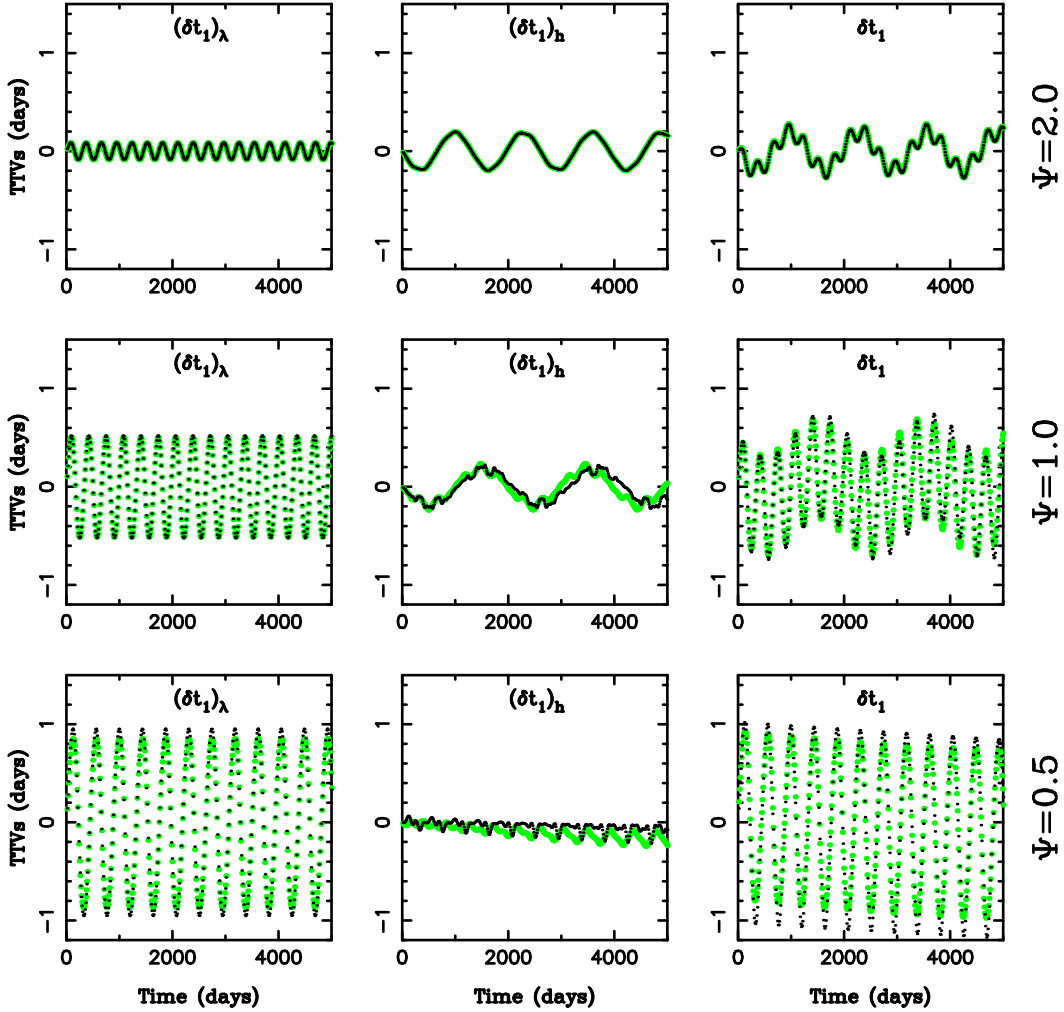}
 \caption{Tests of the validity of the analytic model for different  libration amplitudes. The 
planetary masses and orbital parameters were chosen such that $\delta=2$ in all cases 
shown here. We fixed $m_1=m_2=3\times10^{-4}M_*$, $M_*=1\ M_{\rm Sun}$, $a_1=0.1$ AU, and varied 
$a_2$ near the external 3:2 resonance with the inner planet ($a_2 \simeq 0.131$ AU). The 
eccentricities were adjusted to give $\delta=2$ and a desired initial value $\Psi_0$. 
From top to bottom, the panels show the results for $\Psi_0=2$, $\Psi_0=2.5$ and $\Psi_0=3.5$. 
The stable equilibrium point is located at $\Psi_{\rm eq}=2.24$ for $\delta=2$. The different 
cases shown here thus correspond to the libration amplitudes $A_\Psi=0.25$, 1.24 and 1.74 (from top to 
bottom). The plots show TTVs of the inner planet (the results for the outer planet are 
similar). The green dots were obtained by numerically integrating the differential equations 
corresponding to the resonant Hamiltonian (\ref{ham3}) and (\ref{ham4}). The black lines were obtained from the 
analytic TTV expressions (\ref{ttv}), (\ref{dl}) and a generalization of (\ref{yys2}) derived in 
Section 4.3. From left to right the different panels show the TTVs from $\delta \lambda_j$ and 
$\delta h_j$, and their sum from Eq.~(\ref{ttv}).}
 \label{ttv3}
\end{figure*}

\begin{figure*}
 \epsscale{0.9}
  \plotone{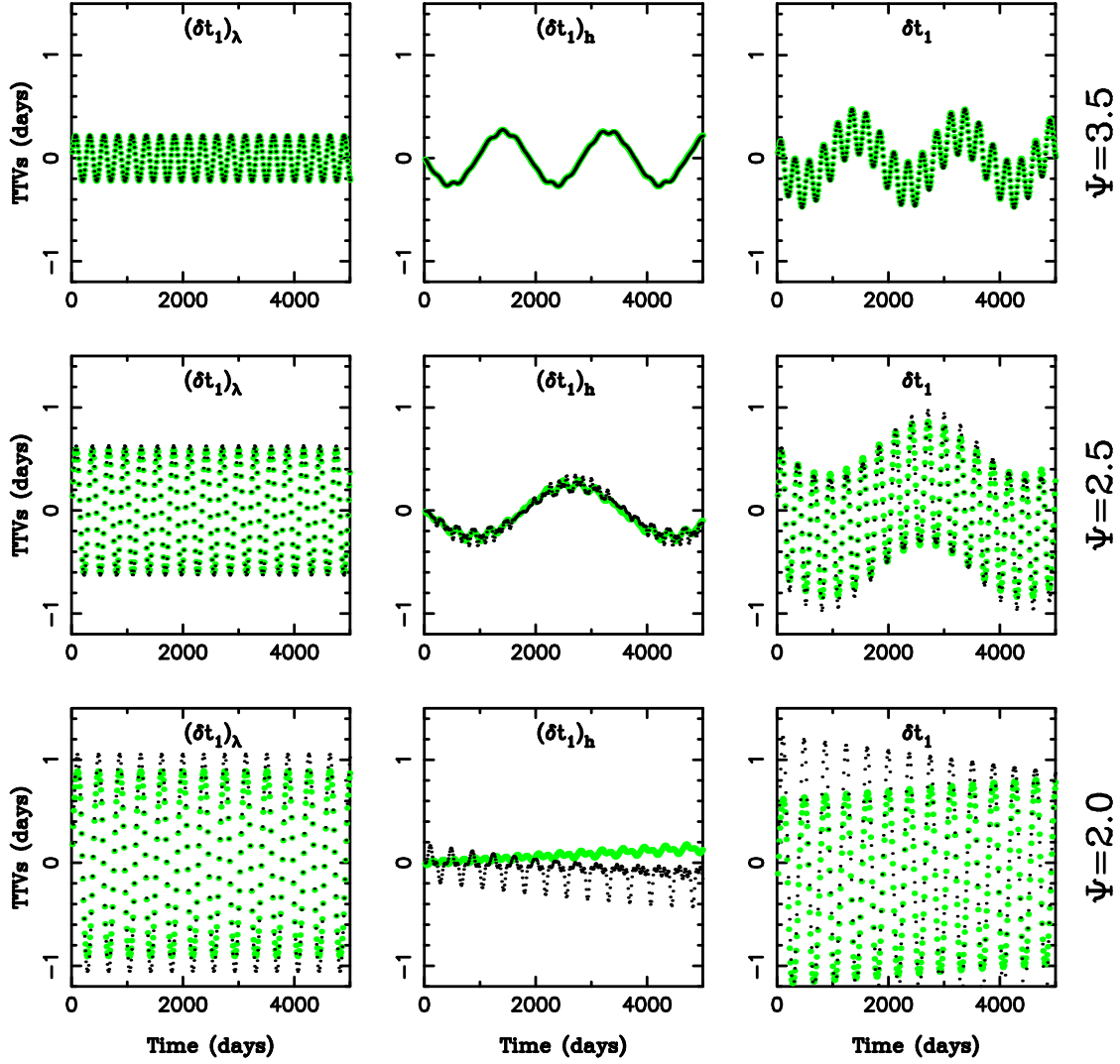}
 \caption{The same as Figure \ref{ttv3} but for $\delta=4$. In this case, the stable equilibrium 
point is at $\Psi_{\rm eq}=4.14$. From top to bottom, different panels correspond to $\Psi_0=3.5$
($A_\Psi=0.67$), $\Psi_0=2.5$ ($A_\Psi=1.17$) and $\Psi_0=2$ ($A_\Psi=1.57$).}
 \label{ttv4}
\end{figure*}

We first test the approximation {\bf V}. To this end we compare the results obtained with code {\bf B}
with the analytic results from method {\bf D}. In {\bf B}, we include the two first-order resonant 
terms and neglect terms that are the second or higher order in planetary eccentricities.\footnote{The same 
comparison method was used to produce Figs. \ref{ttv1} and \ref{ttv2} in Section 2.} The masses and initial
orbits are chosen such that $\delta=2$ (Fig. \ref{ttv3}) or $\delta=4$ (Fig. \ref{ttv4}). The initial orbits 
are then varied to survey different libration amplitudes. These tests show that the analytic method
produces very reliable results for $A_\Psi \lesssim 1$ (top panels in Figs. \ref{ttv3} and \ref{ttv4}). For 
the libration amplitudes much larger than that, our analytic expressions for $u_1$ and $v_1$ in Eq. (\ref{uvapp})
become inaccurate (see Fig. \ref{uvfig}). As a consequence, the analytic approximation of the TTV terms 
from $\delta h_j$ fails (bottom panels in Figs. \ref{ttv3} and \ref{ttv4}). 

Interestingly, however, the analytic approximation of the {\it full} TTV signal is reasonable even if 
$A_\Psi > 1$. This happens because the TTV terms from $\delta \lambda_j$ increase with $A_\Psi$ and become
dominant for large $A_\Psi$. We are able to reproduce these terms correctly, because the analytic formula 
in Eq. (\ref{dl}) has more a general validity than the one that requires a correct approximation of the 
boomerang-shaped trajectories in the $(u_1,v_1)$ plane (Eq. \ref{uvapp}). We therefore conclude that the 
analytic TTV model can be used, with some caution, even if the libration amplitudes are relatively large. 

\begin{figure*}
 \epsscale{0.9}
  \plotone{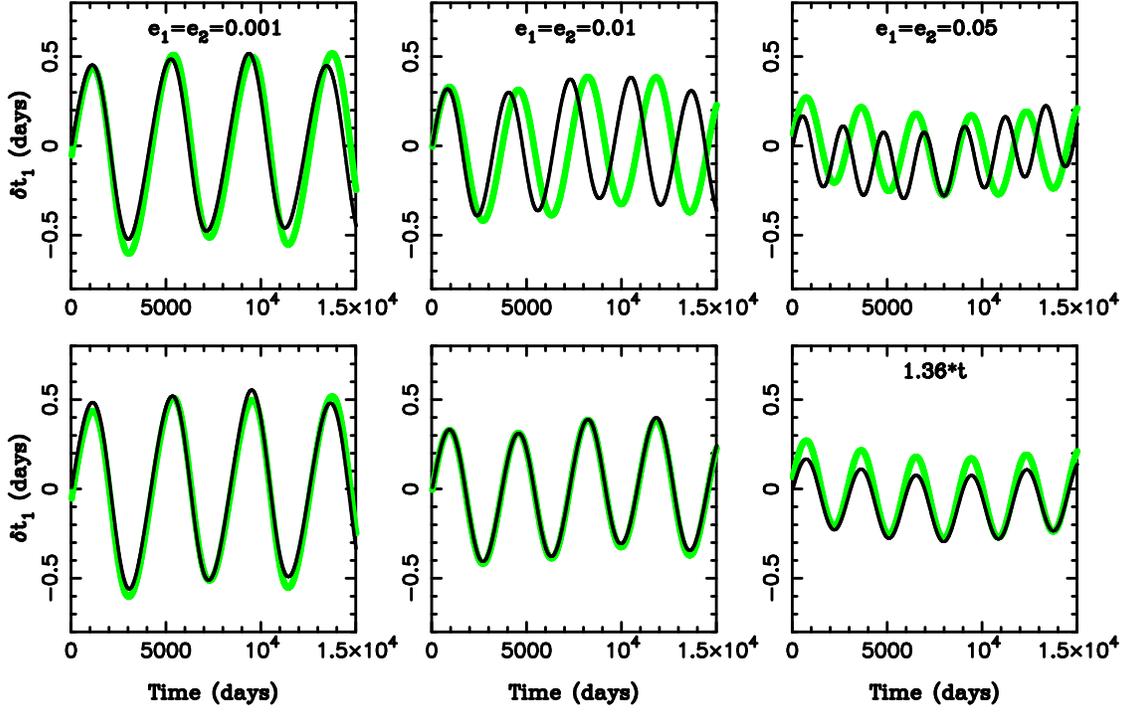}
 \caption{A comparison of TTVs obtained with the analytic model (black lines) and precise $N$-body integrator
(green lines). From left to right, the orbital eccentricities of planets were increased to test the 
validity of the analytic model. We fixed $m_1=m_2=10^{-5}M_*$, $M_*=1\ M_{\rm Sun}$, $a_1=0.1$ AU, and 
$a_2$ near the external 3:2 resonance with the inner planet ($a_2 = 0.13095$ AU). TTVs of the inner planet are 
shown here. The case with $e_1=e_2=0.001$ (left panels) corresponds to $\delta=0.51$ and $\Psi_0=0.007$.
The case with $e_1=e_2=0.01$ (middle panels) corresponds to $\delta=1.25$ and $\Psi_0=0.75$.
The case with $e_1=e_2=0.05$ corresponds to $\delta=19.3$ and $\Psi_0=18.8$. In the two bottom panels 
on the left, we illustrate how a small adjustment of the initial conditions improves the results (see 
the main text for a discussion). In the bottom-right panel, we rescaled time to show that a modest 
adjustment of the frequency can resolve the discrepancy when eccentricities are larger.}
\label{ttv7}
\end{figure*}

\begin{figure*}
 \epsscale{0.9}
  \plotone{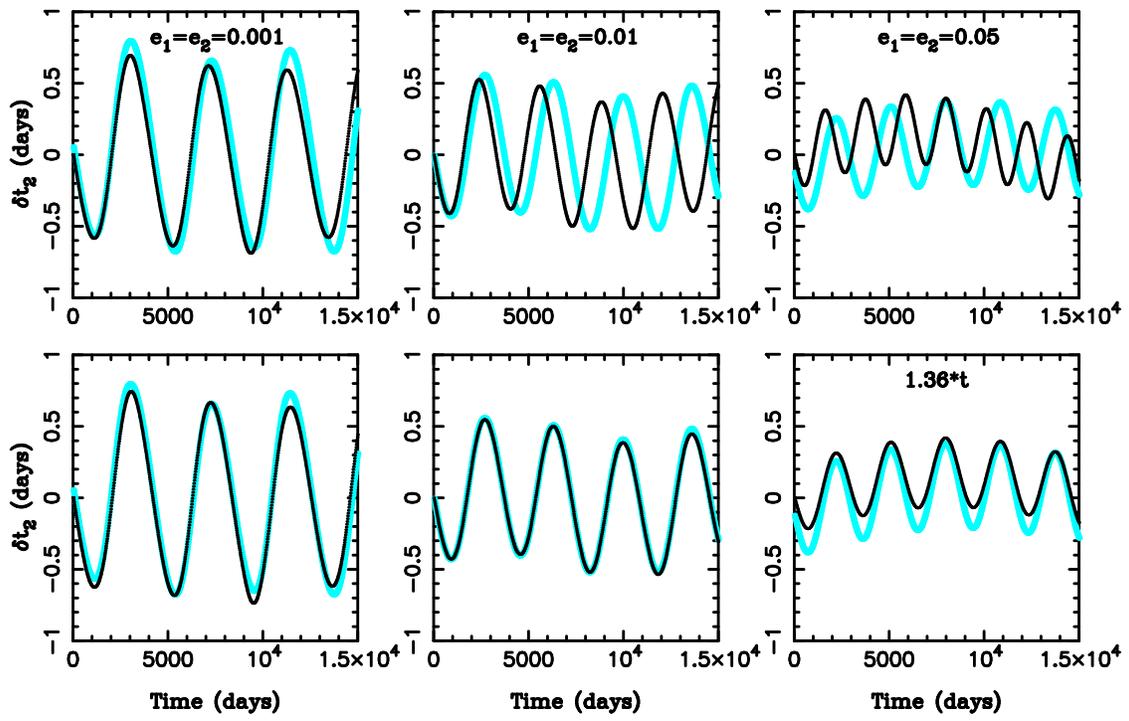}
 \caption{The same as Fig. \ref{ttv7} but for TTVs of the outer planet.}
\label{ttv8}
\end{figure*}

We now turn our attention to the approximations {\bf I, II and III}. We find that the omission of the higher-order
resonant terms in {\bf III} is the most restrictive assumption. To illustrate this, Figures \ref{ttv7} (inner 
planet) and \ref{ttv8} (outer planet) show a comparison of the analytic model with TTVs computed from the 
$N$-body code (method {\bf A} above). Here we set different planetary eccentricities ranging from 
$e_1=e_2=0.001$ (left panels in both figures) to $e_1=e_2=0.05$ (right panels). We find that that the analytic 
model works well for $e_1=e_2=0.001$. 

Already for $e_1=e_2=0.01$, a significant discrepancy appears (see middle panels in Figs. \ref{ttv7} and \ref{ttv8}). 
An important part of the discrepancy, however, is not due to the assumption {\bf III}, but is rather related to the 
choice of initial conditions. Recall that, in addition to the resonant terms, the exact computation of TTVs with 
method {\bf A} also contains the short-periodic harmonics, while the analytic method {\bf D} does not account for these terms. 
This presents a difficulty when choosing the initial conditions in {\bf A} and {\bf D} that are consistent with 
each other. If the same values are adopted in {\bf A} and {\bf D}, the initial semimajor axes in {\bf A} generate 
slightly different values of the mean orbital frequencies than the same initial semimajor axes in {\bf D}. This 
effect then propagates into a difference in the libration frequency $f$. To demostrate this, we surveyed a small
neigborhood of the initial conditions and found that it is always possible, if the eccentricities are sufficiently 
small, to apply a small adjustment such that the difference between the analytical and numerical results vanishes
(left and middle bottom panels in Figs. \ref{ttv7} and \ref{ttv8}). Note that this initial value problem does not seriously 
limit the application of the analytic model to the real data, because it requires only a very small adjustment 
of $a_1$ or $a_2$ (or equivalently $n_1$ and $n_2$), which can easily be absorbed by other parameters. 

Another more fundamental discrepancy appears for $e_1=e_2=0.05$. In this case, the TTV frequency computed from 
the analytic model is nearly 40\% higher than the actual frequency, and the TTV amplitudes are $\simeq$25\% 
smaller than their actual values (right panels in Figs. \ref{ttv7} and \ref{ttv8}). In this case, it is 
not possible to adjust the initial conditions to cancel the difference.
This shows that the assumption {\bf III} starts to fail. 
We confirm this by method {\bf B}, where it becomes apparent that including the second-order resonant terms 
improves model's precision. Still, for $e_1=e_2=0.05$ the amplitude discrepancy is relatively
minor and can be compensated, for example, by a relatively small correction of planetary masses. We therefore find 
that the analytic model is still useful in this case. Our additional tests show that the analytic expressions for 
TTVs are not reliable for eccentricities exceeding $\sim$0.1. [The validity domain in $e$ should be slightly larger 
for distant resonances such as 2:1, and smaller for $k \geq 4$.]   

The analytic model was developed under the assumption of exactly co-planar planetary orbits. This assumption
was used in Section 3.1 to neglect all terms in the Laplacian expansion of the perturbing function that depend 
on inclinations. The model is therefore not expected to be valid if the mutual inclination between orbits,
$I_{\rm mutual}$, is large. We performed various tests of this assumption and found that the analytic model
is reasonably accurate for $I_{\rm mutual}<10^\circ$, but fails to produce reliable results for 
$I_{\rm mutual}\gtrsim 10^\circ$. This should not be a severe limitation of the applicability of the analytic 
results to the multi-transiting planetary systems, because the orbits in these systems are expected to be 
nearly co-planar (e.g, Fang \& Margot 2012).   
\section{Conclusions}
In this work we developed an analytic model for TTVs of a pair of resonant planets, and discussed how
the TTV period and amplitude constrain the masses and orbits of the two planets. The model is strictly 
valid only for small orbital eccentricities ($e<0.1$). It was developed under the assumption of 
co-planar orbits but our tests show that it is valid even if the mutual inclination of orbits is 
not large ($<10^\circ$). 

The resonant TTV signal is expected to contain the harmonics of two basic periods: 
the period of resonant librations and the period of apsidal precession of orbits. The latter is 
expected to be $\sim$5 times longer than the former, and may be difficult to detect with a short 
baseline of the TTV measurements. The libration period is relatively insensitive to the exact 
location of the system parameters in the resonant island, and scales with $(m/M_*)^{-2/3}$.
Its determination from the TTV measurements can therefore help to constrain the planetary masses.  
This is an important difference with respect to the near-resonant case (Lithwick et al. 2012), where
the TTV period is the super period, which is independent of mass. 

The TTV amplitudes, on the other hand, can be used to constrain the ratio of planetary masses $m_1/m_2$. 
Since both the TTV period and amplitude depend on the resonant amplitude $A_\Psi$, some mild degeneracies 
between the mass and orbital parameters are expected, but these degeneracies can be broken by a detection
of higher-order resonant harmonics, which constrain $A_\Psi$, and/or short-periodic (chopping) effects.   
A detailed analysis of this problem and the application of our analytic model to the resonant exoplanets 
(Winn \& Fabrycky 2015) is left for future work.
\acknowledgements
The work of DN was supported by NASA's ADAP program. DV was supported by the Czech Grant Agency 
(grant P209-13-01308S). Slawek Breiter pointed to us the general solution Eq. (\ref{sfmr5}) 
in Whittaker \& Watson (1920). We thank Katherine Deck, Eric Agol and an anonymous reviewer 
for useful comments on the manuscript. 

\appendix

\section{Parameter $\epsilon=b/a$}

The coefficients $C_n$ in Eq. (\ref{sfmr5}) are obtained from the derivatives of $f(\Psi)$ with
$\Psi=\Psi_0$
\begin{eqnarray}
C_1 & = & 2\left[1-2(\Psi_0-\delta)\sqrt{2 \Psi_0}\right]\; , \nonumber \\
C_2 & = & -4\left[2(\Psi_0-\delta)^2+\sqrt{2\Psi_0}\right]\; .
\end{eqnarray} 
The expressions for $C_3$ and $C_4$ are not needed if we set $\psi_0=\pi$ and thus $f_0=0$.
The invariants in Eq. (\ref{sfmr6b}) can be written as
\begin{eqnarray}
g_2 & = & {4 \over 3} \left( h_0^2- {3 \over 2} \delta \right)\; , \nonumber \\
g_3 & = & {1 \over 4} - {8 \over 27} h_0 \left(h_0^2 - {9 \over 4} \delta \right)\; ,
\end{eqnarray}
where we denoted $h_0=-(\Psi_0-\delta)^2+\sqrt{2 \Psi_0}$. The determinant $\Delta$ becomes
\begin{equation}
\Delta = - {27 \over 16} - 8 \delta^3 + 4 h_0 
\left( h_0^2 + \delta^2 h_0 -{9 \over 4}\delta\right)\; .
\end{equation}
If we define 
\begin{equation}
F={1 \over 2} \left( g_3 + {1 \over 3} \sqrt{- {\Delta \over 3}} \right)^{1/3}\; , 
\end{equation}
then the three roots of the cubic equation can be obtained from
\begin{eqnarray} 
\alpha & = & - {1 \over 2} \left ( F + { g_2 \over 12 F} \right)\; , \nonumber \\
\beta & = & {1 \over 2 \sqrt{3}} \left ( F - { g_2 \over 12 F} \right)\; , \nonumber \\
e_2 & = & F + {g_2 \over 12F} = -2 \alpha  \; .
\end{eqnarray}

In the equilibrium point, $\Psi_0=\Psi_{\rm eq}$, we have that $\Delta=0$ and thus 
$F=g_3^{1/3}/2$. It follows that $\beta=0$ and $\alpha=-F=-g_3^{1/3}/2=C_2/24<0$.
Therefore, $b = 0$ and $a = -C_2/4$, where $a$ and $b$ are defined in the main 
text. Figure \ref{eps} shows $\epsilon=b/a$ for $\delta=3$. It is zero at the 
equilibrium point and increases to $\epsilon \simeq 0.2$ at the separatrix. 
Higher values of $\delta$ lead to smaller values of $\epsilon$. For $\delta 
\simeq 1$, on the other hand, $\epsilon$ can be as large as 0.6 near the 
separatrix.

\section{Expressions for $Y_j$}

In Section 3 we omitted to explain one important issue
that becomes apparent if the degree of freedom related to 
$\theta=k\lambda_2-(k-1)\lambda_1$ is treated separately from those related 
to $\varpi_1$ and $\varpi_2$ (as it was done in BM13). To explain this issue we first
define in direct correspondence to Eq. (\ref{ham10b})
\begin{eqnarray}
  X_1 & \!\!\!=\!\!\! & \sqrt{2\Gamma_1}\cos\gamma_1\; , \quad
   Y_1= \sqrt{2\Gamma_1}\sin\gamma_1 \; , \nonumber \\
  X_2 & \!\!\!=\!\!\! & \sqrt{2\Gamma_2}\cos\gamma_2\; , \quad
   Y_2= \sqrt{2\Gamma_2}\sin\gamma_2 \; , \label{gamma} 
\end{eqnarray}
where $\gamma_j=-\varpi_j$. Second, we perform a transformation (see 
Section 3.4) to the new variables $(V_1,V_2;U_1,U_2)$ 
\begin{eqnarray}
  U_1 & \!\!\!=\!\!\! & \frac{AX_1+BX_2}{\sqrt{A^2+B^2}}\; , \quad
   V_1= \frac{AY_1+BY_2}{\sqrt{A^2+B^2}} \; ,  \nonumber \\
  U_2 & \!\!\!=\!\!\! & \frac{BX_1-AX_2}{\sqrt{A^2+B^2}}\; , \quad
   V_2= \frac{BY_1-AY_2}{\sqrt{A^2+B^2}} \; . \label{bigredt}
\end{eqnarray}
And last, we introduce new polar variables $(\zeta_1,\zeta_2;\Phi_1,\Phi_2)$ 
such that \begin{eqnarray}
  U_1 & \!\!\!=\!\!\! & \sqrt{2\Phi_1}\cos\zeta_1\; , \quad
   V_1= \sqrt{2\Phi_1}\sin\zeta_1 \; , \nonumber \\
  U_2 & \!\!\!=\!\!\! & \sqrt{2\Phi_2}\cos\zeta_2\; , \quad
   V_2= \sqrt{2\Phi_2}\sin\zeta_2 \; . \label{zeta} 
\end{eqnarray}
The inverse transformation to (\ref{bigredt}) is  
\begin{eqnarray}
  X_1 & \!\!\!=\!\!\! & \frac{AU_1+BU_2}{\sqrt{A^2+B^2}}\; , \quad
   Y_1= \frac{AV_1+BV_2}{\sqrt{A^2+B^2}} \; , \nonumber \\
  X_2 & \!\!\!=\!\!\! & \frac{BU_1-AU_2}{\sqrt{A^2+B^2}}\; , \quad
   Y_2= \frac{BV_1-AV_2}{\sqrt{A^2+B^2}} \label{inverse}\; .
\end{eqnarray}

With these definitions, it is straightforward to show that $\phi_j=\theta+\zeta_j$,
where $\phi_j$ are the original angles defined in (\ref{ham14b}), $\zeta_1 
= {\rm arg}(A Z_1 + B Z_2)$ and $\zeta_2 = {\rm arg}(B Z_1 - A Z_2)$, where
$Z_j=X_j+\imath Y_j = \sqrt{2 \Gamma_j} \exp \imath \gamma_j$. We thus find 
that $V_1=v_1 \cos \theta - u_1 \sin \theta$. When substituted into (\ref{inverse}),
we obtain Eq. (\ref{ys}) in the main text. In addition, it can be shown that
${\rm d}\zeta_2/{\rm d}t= {\rm d}\phi_2/{\rm d}t-{\rm d}\theta/{\rm d}t=0$. The 
angle $\zeta_2$ is therefore constant. Consequently, since $\Phi_2={\rm const.}$ 
as well, both $U_2$ and $V_2$ are constants of motion. This result is used in 
Section 4, where $V_2$ in Eq. (\ref{ys}) does not contribute to TTVs. 

\section{Integral $\int \Psi {\rm d}\tau$}

The integral in Eq. (\ref{iscale}) with $\Psi(\tau)$ from Eq. (\ref{simple}) 
admits the following exact solution
\begin{equation}
 \int \Psi\left(\tau\right)\, {\rm d}\tau =
  \left(\Psi_0- \frac{C_1}{4a\epsilon}\right)\tau +
  \frac{C_1(1+\epsilon)}{8a\epsilon\sqrt{\gamma}}\int \frac{{\rm d}u}{1+\epsilon\,
  {\rm cn}(u,k)} \; ,
\end{equation}
where $\epsilon=b/a$ and $\gamma$ are defined in Section 4.1. From Byrd \& Friedman 
(1971) (BF 341.03) we have
\begin{equation}
 \int \frac{{\rm d}u}{1+\epsilon\, {\rm cn}(u,k)} =
  \frac{1}{1-\epsilon^2} \left[\Pi\left(\varphi,n,k\right)-
  \epsilon\, C\, {\rm atan}\left(\frac{{\rm sd}\left(u,k\right)}{C}
  \right)\right]
  \; ,
\end{equation}
where $\Pi\left(\varphi,n,k\right)$ is the Legendre elliptic integral of the 
third kind, $\varphi={\rm am}\, u$ is the Jacobi amplitude, and ${\rm sd}(u,k) 
= {\rm sn}(u,k)/{\rm dn}(u,k)$. The constants $n$ a $C$ are
\begin{eqnarray}
 n & = & \frac{\epsilon^2}{\epsilon^2-1} \; \\
 C & = & \sqrt{\frac{1-\epsilon^2}{k^2+\epsilon^2\,k'^2}} \; .
\end{eqnarray}
Note that Byrd \& Friedman (1971) use a different notation for the coefficient $n$ 
then Press et~al. (2007). To use the numerical subroutines from Press et~al. (2007), 
$n=\epsilon^2/(1-\epsilon^2)$.

\clearpage
\begin{table}
\centering
{
\begin{tabular}{lrrr}
\hline \hline
res.                  & $\alpha_{\rm res}$  & $f_1$           & $f_2$  \\    
\hline
2:1    & 0.630    & -1.190   & 0.428   \\     
3:2    & 0.763    & -2.025   & 2.484   \\
4:3    & 0.825    & -2.840   & 3.283   \\
5:4    & 0.862    & -3.650   & 4.084   \\
6:5    & 0.886    & -4.456   & 4.885   \\
7:6    & 0.902    & -5.261   & 5.686   \\                          
\hline \hline
\end{tabular}
}
\caption{The coefficients $f_1$ and $f_2$ for different resonances. In the second 
column, we report the semimajor axis ratio for an exact resonance. The coefficient 
values are given for $\alpha=a_1/a_2=\alpha_{\rm res}$.}
\end{table}

\clearpage

\clearpage

\clearpage

\clearpage

\clearpage

\clearpage

\clearpage

\clearpage

\clearpage

\clearpage

\clearpage

\clearpage

\clearpage

\clearpage



\clearpage

\clearpage

\end{document}